\documentclass[twocolumn,superscriptaddress,amssymb, aps, prc,nobibnotes,floatfix,showpacs, nofootinbib]{revtex4-2}
\usepackage{amsmath}
\usepackage{amssymb}
\usepackage{graphicx}
\usepackage{color}
\usepackage{float}
\usepackage{longtable}
\usepackage{afterpage}
\usepackage{rotating}
\usepackage{mathptmx}

\usepackage[colorlinks=true, citecolor=blue,linkcolor=blue]{hyperref}

\begin{document}

\title{\Large Coexisting single-particle excitations and octupole correlations in transitional nucleus $\mathbf{^{217}Ra}$}%

\author{Madhu}
\author{A. Y. Deo}
\email[Corresponding author:]{ajay.deo@ph.iitr.ac.in}
\author{Khamosh Yadav}
\author{Dhananjaya Sahoo}
\affiliation{Department of Physics, Indian Institute of
Technology Roorkee, Roorkee 247667, India}
\author{Y. Y. Wang}
\affiliation{Mathematics and Physics Department, North China Electric Power University,
Beijing 102206, China}
\author{Y. K. Wang}
\affiliation{State Key Laboratory of Nuclear Physics and Technology, School of Physics, Peking University, Beijing 100871, China}
\author{J. Meng}
\affiliation{State Key Laboratory of Nuclear Physics and Technology, School of Physics, Peking University, Beijing 100871, China}
\affiliation{Yukawa Institute for Theoretical Physics, Kyoto University, Kyoto 606-8502, Japan}
\author{Saket Suman}
\author{S. K. Tandel}
\affiliation{School of Physical Sciences, UM-DAE Centre
for Excellence in Basic Sciences,
University of Mumbai, Mumbai 400098, India}
\author{A. Sharma}
\affiliation{Department of Physics, Himachal Pradesh University, Shimla 171005, INDIA}
\author{I. Ahmed}
\author{K. Katre}
\author{K. Rojeeta Devi}
\author{S. Dutt}
\author{S. Kumar}
\author{Yashraj}
\author{S. Muralithar}
\author{R. P. Singh}
\affiliation{Inter University Accelerator Centre,
Aruna Asaf Ali Marg, New Delhi 110067, India}

\date{\today}%
\begin{abstract}
{
The level structure of the transitional nucleus $\mathrm{^{217}Ra}$ has
been extended with the addition of around 20 new transitions.
The discrepancies between the placements of several transitions 
reported in the earlier studies are resolved. The newly-established
negative-parity sequence at low excitation energies hints at the
expected parity-doublet structures in this nucleus. 
The properties of the observed simplex bands are compared with that
of similar bands in neighboring nuclei. Since the presence of
parity-doublet structures reflects octupole correlations, theoretical
calculations using reflection-asymmetric triaxial particle rotor
model (RAT-PRM) have been performed. A comparison of the observed
features of the simplex bands with the predictions of the RAT-PRM
calculations suggests that $\mathrm{^{217}Ra}$ exhibits an intermediate
behavior between the extremes of spherical and
octupole-deformed nuclei. The termination of the
simplex bands at intermediate energies and the structures lying above reflect
the dominance of the single-particle excitations at higher excitation energies.
}
\end{abstract}

\maketitle

\section{Introduction}
\label{Section_1}
The atomic nucleus is a many-body quantal system in which the
number of valence nucleons outside the shell closure and the
interactions between them mainly govern shape and properties
of the system. At the beginning of several major shells, the
presence of an opposite-parity intruder orbital in the vicinity
of normal-parity orbitals with $\Delta$\textit{j} = $\Delta$\textit{l} = 3 
gives rise to long-range octupole correlations between the
valence nucleons \cite{Ahmed_ButlerReview}. Such orbitals are
present near the Fermi surface in nuclei with \textit{Z} or
\textit{N} $\simeq$ 34, 56, 88 and 134 where the fingerprints of
dynamical or stable reflection asymmetric shapes have been observed
\cite{Ahmed_ButlerReview, Butler_NazarewiczReview,SCP_book}.
The advancements in the experimental facilities with the availability
of radioactive ion beams and targets have facilitated the measurements
of the \textit{E}3 transition strengths using Coulomb excitation
techniques \cite{226Ra,224Ra_Nature,222Ra_PRL,144Ba,146Ba}.
The deduced electric octupole moments, which are a direct measure of the
strength of octupole correlations, in the Ra isotopes with 134
$\leq N \leq$ 138 ($\mathrm{^{222}Ra}$, $\mathrm{^{224}Ra}$, $\mathrm{^{226}Ra}$)
have confirmed the stable octupole deformation in these nuclei
\cite{226Ra,224Ra_Nature,222Ra_PRL,Butler_RoyalSociety_Review}.
In terms of the level structures, static quadrupole and octupole
deformation in the even-even nuclei is characterized by $\Delta$\textit{I} = 2
sequences of opposite parities connected by enhanced \textit{E}1 transitions. 
The excitation energies of the states of these $\Delta$\textit{I} = 2 
bands follow the \textit{I}(\textit{I}+1) dependence as
expected for rotational bands. Also, the negative-parity sequence
is expected to be perfectly interleaved with the positive-parity
ground state band in the case of static octupole deformation.
In addition, the typical \textit{B}(\textit{E}1)/\textit{B}({\textit{E}2)
ratios are of the order of $\sim$ 10$^{-6}$ $\mathrm{fm^{-2}}$
for such nuclei \cite{Ahmed_ButlerReview, Butler_NazarewiczReview,SCP_book}.
For the case of odd-\textit{A} and doubly-odd nuclei where the
odd nucleon tilts the total angular momentum with respect
to the normal to the symmetry plane, the simplex symmetry
is broken \cite{Frauendorf}. In such nuclei, parity-doublet structures
(almost degenerate states of the same spin but
opposite parity) are expected \cite{Ahmed_ButlerReview, Frauendorf}.

As one moves away from the region of stable octupole deformation
towards the shell closure, a transitional region is encountered
where the interplay between both the single-particle and collective
degrees of freedom results in complex structures and the properties
mentioned above become less pronounced. The recent high-spin
investigations in $\it {N}$ = 130 isotones
($\mathrm{^{218}Ra}$ and $\mathrm{^{220}Th}$) and $\mathrm{^{219}Th}$ 
report near-constant transition energies and spin-dependent staggering
of \textit{B}(\textit{E}1)/\textit{B}({\textit{E}2) ratios in these
nuclei \cite{220Th_TidalWave, N130_TidalWave, 219Th_Reviol}.
The almost constant level spacings and angular velocity over a
spin range imply that the nucleus does not gain angular momentum
by rotating faster. Thus, a new approach in terms of a reflection
asymmetric tidal wave running over the nuclear surface has been
proposed to interpret the observed properties in these nuclei \cite{Frauendorf_TidalWave}.

It has been noted that the properties of the level structures of nuclei
in the transitional region exhibit a significant dependence
on the nucleon number. The low-lying states in $\mathrm{^{216}Ra}$,
with two valence neutrons outside the shell closure, exhibit
decreasing level spacing with spin as expected for single-particle
$\nu g_{9/2}^{2}$ excitations \cite{216Ra_Lonnorth, 216Ra_Muralithar}. 
On the other hand, spectroscopic studies in $\mathrm{^{218}Ra}$
(\textit{N} = 130) revealed irregular sequences of opposite parity
connected by enhanced \textit{E}1 transitions which, in turn, reflect
the dominance of octupole correlations over quadrupole deformation in
this nucleus \cite{218Ra_Schulz}. In $\mathrm{^{217}Ra}$, which lies
in between the above mentioned isotopes, three sequences of \textit{E}2
transitions are reported at low excitation energies
out of which two are connected by \textit{E}1 transitions
\cite{217Ra_Roy,217Ra_Sugawara, 217Ra_Sugawara_RIKEN, 217Ra_Muralithar}.
Although the alternating parity sequence was qualitatively
understood in terms of single-particle neutron excitations,
the observed near linear spin dependence of excitation
energies and large \textit{B}(\textit{E}1)/\textit{B}({\textit{E}2)
ratios ($\sim$ 10$^{-6}$ $\mathrm{fm^{-2}}$) hint at the 
contribution of collective degrees of freedom in this nucleus.
Thus, the disparate level structure in $\mathrm{^{217}Ra}$ 
with respect to its neighboring even-even isotopes requires
a detailed spectroscopic study of this nucleus in order to understand
how the dominance of underlying degrees of freedom evolves
in the transitional region.

Prior to the present work, high-spin studies in $\mathrm{^{217}Ra}$
were carried out by several groups in which levels up to 5 MeV excitation
energy were established using $\gamma$-$\gamma$ coincidences,
angular distribution measurements and conversion-electron spectroscopy
\cite{217Ra_Roy,217Ra_Sugawara,217Ra_Sugawara_RIKEN,217Ra_Muralithar}. Although the
reported low-lying sequences are consistent with each other, the
placement and ordering of several transitions were found to be at
variance in these studies and need to be revisited. Also, the
spin-parities of the excited states above 2.4 MeV energy
were reported to be tentative in these studies.
It may be noted that the low-lying level structure in
$\mathrm{^{217}Ra}$ is observed to be similar to its
\textit{N} = 129 isotones \cite{216Fr,218Ac_Debray,219Th_Reviol}. 
However, in all the \textit{N} = 129 isotones with 87 $\leq Z \leq$ 90 nuclei,
except $\mathrm{^{217}Ra}$, parity-doublet structures have been reported.
Therefore, it is imperative to search for similar structures expected
in $\mathrm{^{217}Ra}$ and understand the observed properties with the
help of the relevant nuclear models. Thus, an extensive $\gamma$-ray
spectroscopy study of $\mathrm{^{217}Ra}$ has been carried out
using the heavy-ion fusion reaction mechanism to investigate
high-spin states and the evolution of structure with angular
momentum in this nucleus. In the present work, we report an
extended level scheme of $\mathrm{^{217}Ra}$ with the addition of
around 20 new transitions. We provide the first evidence of a
new negative-parity sequence which constitutes the simplex
partner band of the already reported ground state alternating
parity sequence. Also, the observed level structures in $\mathrm{^{217}Ra}$
reflect the coexistence of single-particle and collective
excitations in this transitional nucleus.

 \begin{figure*}[ht!]
\begin{centering}
  \includegraphics*[angle = 270, width=\textwidth]{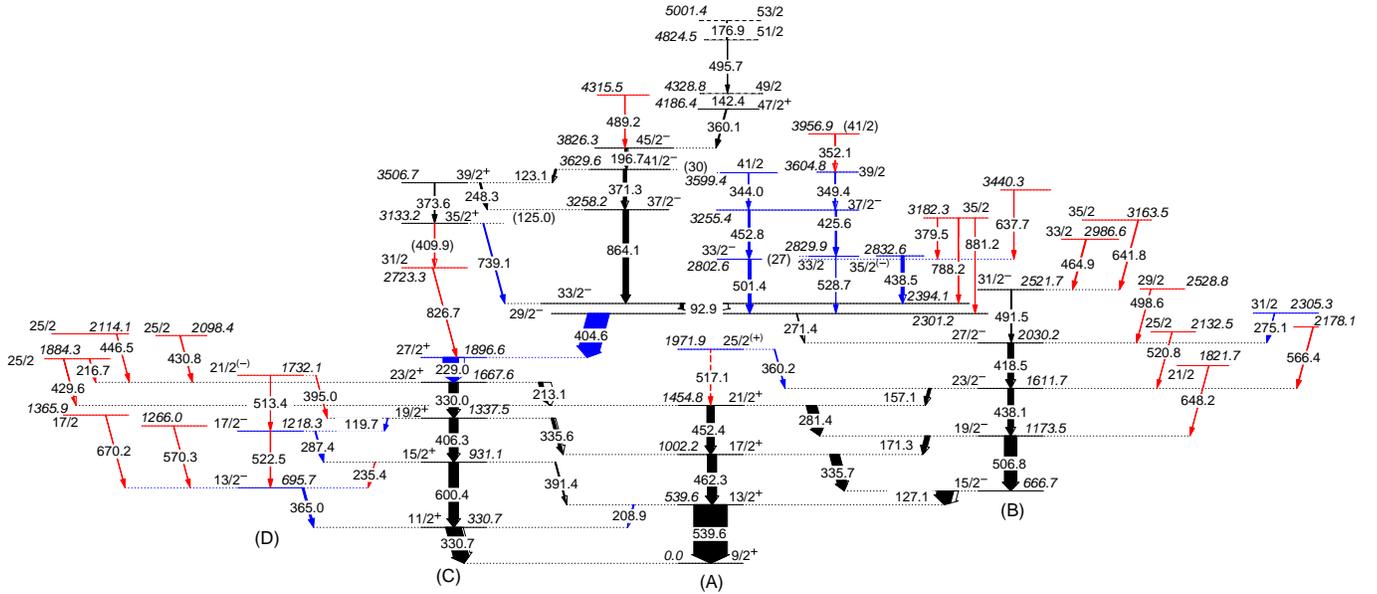}%
\end{centering}
\caption{\label{LS}The level scheme of $^{217}$Ra
established in the present work. The widths of the closed and open
areas of the arrows correspond to the intensity of the $\gamma$ rays
and conversion electrons, respectively. The newly identified
$\gamma$ rays and levels are presented in red color. The transitions
whose placements were inconsistent in different studies
\cite{217Ra_Roy, 217Ra_Sugawara_RIKEN, 217Ra_Muralithar}
or uncertain in Ref. \cite{217Ra_NDS} and have been confirmed 
in the present work are shown in blue color. In addition,
the earlier reported $\gamma$ rays whose ordering has been changed
are also shown in blue color.
}
\end{figure*}


\section{Experimental Details}
\label{Section_2}
High-spin states in $\mathrm{^{217}Ra}$ were populated using the
$\mathrm{^{208}Pb (^{12}C, 3n){^{217}Ra}}$ reaction.
The $^{12}$C beam with 68-, 72- and 80 MeV
energies was provided by the 15UD Pelletron accelerator at
the Inter-University Accelerator Centre, New Delhi.
The target consisted of an isotopically enriched (99\%) and
self-supporting $^{208}$Pb foil of $\sim$ 9 mg/{cm}${^2}$ thickness.
The $\gamma$ rays deexciting the residual nuclei were detected using
the Indian National Gamma Array (INGA) \cite{INGA} comprising of 13
Compton suppressed clover detectors and one Low-Energy Photon
Spectrometer (LEPS). The detectors were placed in five rings
at $32^{\circ}$, $57^{\circ}$, $90^{\circ}$, $123^{\circ}$
 and $148^{\circ}$ with respect to the beam direction.
The raw data were collected using a VME based data acquisition
system \cite{VME}, which were then written to a disk in a ROOT
\cite{ROOT} Tree format. The energy and efficiency calibration
were performed using a standard $\mathrm{^{152}Eu}$ source. 
The calibrated data were written into another ROOT Tree and
various histograms were generated for further analysis. 
It may be noted that the two- and higher fold
coincidence data collected at the three beam energies were
combined in order to determine the coincidence relationships between
the $\gamma$ rays, multipolarities of the transitions and
lifetime of isomeric states. In addition, single-fold data were
also acquired to obtain the intensity information for the strong and
non-contaminated $\gamma$ rays.

The level scheme was constructed on the basis of coincidence relationships
between the observed $\gamma$ rays and the intensity measurements.
In order to establish the coincidences between the
$\gamma$ rays, the transitions which were detected within 100 ns 
with respect to each other were used to construct symmetric
2- and 3-dimensional prompt histograms.
Several $\textit{early - delayed}$ $\gamma$-$\gamma$ matrices within
different time windows ($\Delta T$) were generated to search for
isomeric states. For such histograms, the transitions which
precede (succeed) the coincident $\gamma$ rays within a certain
time interval ($\Delta T$) are placed on the
$\it{early}$ ($\it{delayed}$) axis. In the present work,
the half-life of the metastable state was extracted using the
centroid-shift analysis. It may be noted the lifetimes greater
than one-twentieth of the FWHM of prompt distribution can be extracted
using the centroid-shift technique \cite{CShift}. For the current detection
setup, the FWHM was found to be around 50 ns for $\gamma$ rays of
$\sim$ 500 keV energy. For the centroid-shift analysis, histograms
of time difference ($\Delta T$) between the $\gamma$ rays feeding
and deexciting the state were generated.

The spin and parities of the states in the level scheme were assigned
on the basis of the multipolarities of the $\gamma$ rays.
Since the intensities of the $\gamma$ rays originating from the
oriented nuclear states exhibit a particular angular dependence based 
on the multipole order of the transitions, the ratios of directional
correlations of oriented states ($R_\textrm{DCO}$) \cite{RDCO} were
measured. An asymmetric $\gamma$-$\gamma$ matrix
was generated with the transitions detected at $32^{\circ}$
and $148^{\circ}$ ($\theta_1$) on the \textit{X} axis and the coincident
transitions detected at $90^{\circ}$ ($\theta_2$) on the \textit{Y} axis.
The $R_\textrm{DCO}$ value for a given transition is defined as,
 \begin{equation}
    R_\textrm{DCO}\ (\gamma_{1}) = \frac{I_{\gamma_{1}}\ at\ \theta_{1} \arrowvert gate\ on\ \gamma_{2}\  at\ \theta_2}{I_{\gamma_{1}}\ at\ \theta_{2}\arrowvert gate\ on\ \gamma_{2}\ at\ \theta_1}
 \end{equation}

It was found that for a stretched quadrupole (dipole) gating transition,
the $R_\textrm{DCO}$ value for a quadrupole (dipole) transition
is 1.0 (1.0) whereas the $R_\textrm{DCO}$ value for a dipole (quadrupole)
$\gamma$ ray is 0.5 (2.0). Since the $R_\textrm{DCO}$ values for the
electric and magnetic transitions of a given multipole order are the same,
polarization measurements were performed in order to ascertain
the parities of the states.

The polarization asymmetry ($\Delta_\textrm{asym}$)
\cite{PDCO, PDCO_2} for a transition of interest is defined as
 
 \begin{equation}
  \Delta_\textrm{asym} = \frac{a(E_\gamma)N_\perp-N_\parallel}{a(E_\gamma)N_\perp+N_\parallel}
 \end{equation}
where N$_\perp$ (N$_\parallel$) denotes the number of
Compton-scattered $\gamma$ rays detected in the direction
perpendicular (parallel) with respect to the reaction plane.
  
The inherent geometrical asymmetry [$a(E_{\gamma}$)] in the polarization
measurements is taken into account by determining the ratio of the
parallel to perpendicular counts ($\frac{N_\parallel}{N_\perp}$)
for unpolarized gamma rays as a function of energy. For the present
detection setup, the value of $a(E_{\gamma}$) was found to be 1.02(2).
The positive value of polarization asymmetry indicates the electric 
nature of the transition of interest, whereas the magnetic transitions
have negative $\Delta_\textrm{asym}$. For the transitions where the
measured $\Delta_\textrm{asym}$ values are near zero and/or
$R_\textrm{DCO}$ measurements suggests mixed nature, information on the
mixing of different multipoles is required. In order to determine the
mixing ratios ($\delta$), theoretical $R_\textrm{DCO}$ and linear
polarization (\textit{P}) values were compared with the experimental
data points. The theoretical values of $R_\textrm{DCO}$ and linear
polarization (\textit{P}) were calculated using the formalism
given in Refs. \cite{Theoretical_AD1,Theoretical_AD2,PolSens_Palit}. The value of
spin alignment parameter was adopted to be 0.3 in the present calculations.

The linear polarization is related to the measured polarization asymmetry as
 
 \begin{equation}
  P = \frac{\Delta_\textrm{asym}}{Q(E_\gamma)}
 \end{equation}
 
 The polarization sensitivity $[Q(E_\gamma)]$ \cite{PDCO_2,PolSens_Palit},
 which depends on the
 geometry of the polarimeter, varies as a function of energy as
 per the following relation
 \begin{equation}
  Q (E_\gamma) = (CE_\gamma+D)Q_0(E_\gamma)
 \end{equation}
 Here, $Q_0$($E_\gamma$) is the polarization sensitivity for a
 point absorber and scatterer and is defined as
 \begin{equation}
 Q_0(E_\gamma) = \frac{\alpha+1}{\alpha^2+\alpha+1}
 \end{equation}
 with $\alpha$ = $E_\gamma$/m$_{e}$c$^2$, where $E_\gamma$ is energy
 of the incident $\gamma$-ray photon and m$_{e}$c$^2$ is the
 rest-mass energy of the electron. The coefficients C and D for the
 polarization sensitivity were taken from Ref. \cite{215Fr_KY}.

\section{Results}

\begin{figure}[b]
 \begin{center}
  \includegraphics*[width=\linewidth]{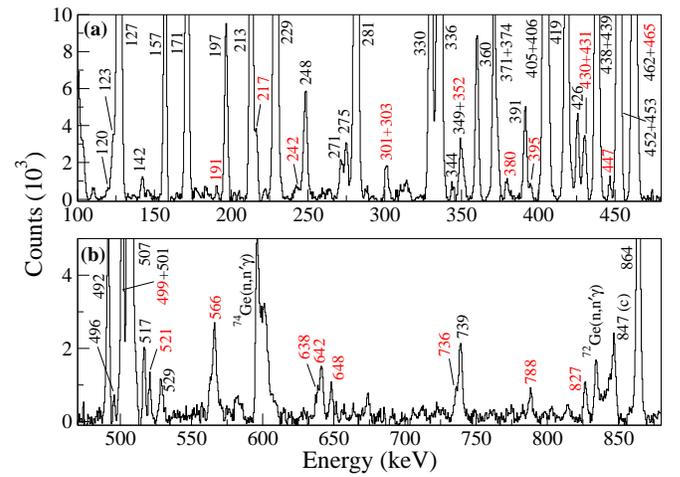}
  
   \caption{\label{g540} Coincidence $\gamma$-ray spectra illustrating
   transitions in the gate of the 540-keV $\gamma$ ray. The new
   transitions are labeled with red color. The
   contaminant line marked with (c) is from the reaction of the
   beam with the Fe target frame.}
 \end{center}
\end{figure} 


Figure \ref{LS} shows the level scheme of $\mathrm{^{217}Ra}$
established in the present study. Apart from a few modifications,
the proposed level structure is found to be mostly consistent
with that reported in Refs. \cite{217Ra_Roy, 217Ra_Sugawara_RIKEN}.
In addition to the earlier reported transitions, around 20 new
$\gamma$ rays have been placed in the level scheme on the
basis of coincidence relationships and intensity measurements.
Also, the placements of several known transitions, which were found
to be inconsistent in the earlier studies, have been confirmed in
the present work. The multipolarities of the observed $\gamma$ rays,
wherever possible, were determined using the $R_\textrm{DCO}$ and
linear-polarization measurements. Table \ref{Table I} lists the 
energies, relative intensities (I$_{\gamma}$), $R_\textrm{DCO}$ and
polarization values of the observed $\gamma$ rays along with the
spin-parities and excitation energies of the states.
Furthermore, the established level scheme was investigated
to search for isomers using $\textit{early-delayed}$ $\gamma$-$\gamma$
matrices. The present analysis does not indicate the presence of any
longer-lived [$\tau >$ 2.5 ns] state other than
the previously reported isomer with $T_{1/2}$ = 4.62(6) ns
at an excitation energy of 2.4 MeV. The detailed results are
discussed in the following subsections.

\subsection{Revisiting the earlier reported level structures}

\begin{figure}[b]
 \begin{center}
  \includegraphics*[width=\linewidth]{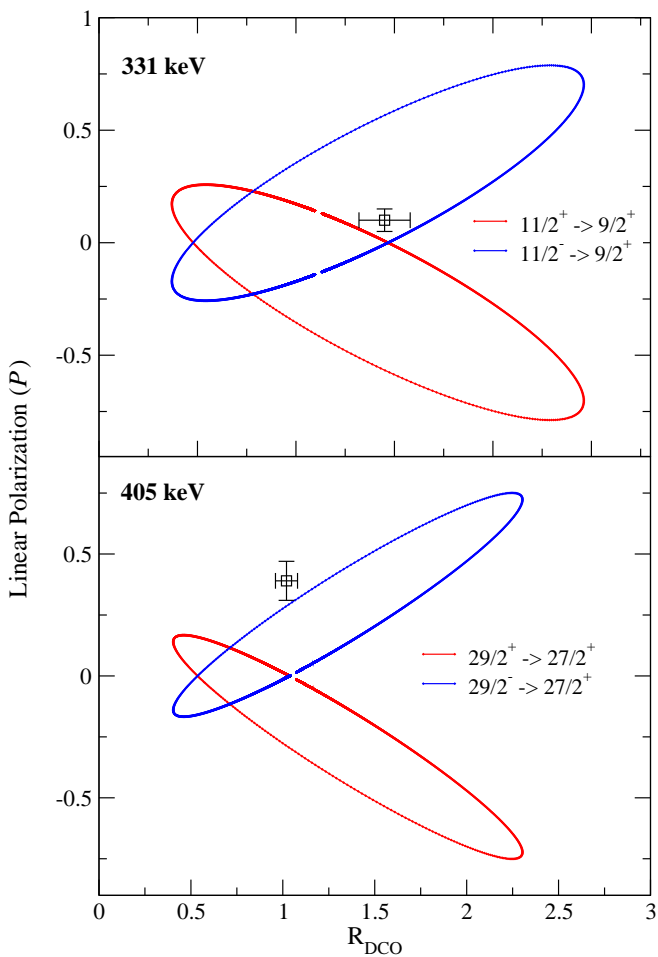}
  
\caption{\label{mixRatio_331} Theoretical contours illustrating the
calculated $R_\textrm{DCO}$ and linear-polarization values (\textit{P})
as a function of mixing ratio ($\delta$), along with the experimental
data points, for the 331- and 405-keV transitions.}
 \end{center}
\end{figure}

Figure \ref{g540} illustrates the $\gamma$ rays in coincidence
with the 540-keV transition, which directly feeds the ground state.
In addition to the several new transitions, most of the $\gamma$ rays
reported in the previous studies have been identified in this
coincidence spectrum. Furthermore, it was observed that the deduced
level structures from the present data are consistent with the reported
ground-state ``A'' and negative-parity ``B'' sequences
\cite{217Ra_Roy,217Ra_Sugawara_RIKEN, 217Ra_Muralithar}.
The level scheme of $\mathrm{^{217}Ra}$ is particularly complex
due to presence of several doublets in this nucleus. The 331-keV
$\gamma$ ray, which is one of the two transitions feeding the 
ground state, is also a doublet and is observed to be in coincidence
with its doublet counterpart 330-keV transition. Thus, in order to
determine the multipolarity of the 331-keV transition, the $\gamma$ rays 
in a parallel decay path with respect to the 330-keV $\gamma$ ray
were used as the gating transitions. The measured $R_\textrm{DCO}$ and
polarization values (see Table \ref{Table I}) indicate the mixed
dipole + quadrupole (\textit{E}1+\textit{M}2 or \textit{M}1+\textit{E}2)
nature of the 331-keV transition. Therefore, in order to determine the
true character of the transition, mixing ratio calculations were performed.
As discussed in Section \ref{Section_2}, the mixing ratio for a transition
of interest is determined by comparing the theoretical and experimental
$R_\textrm{DCO}$ and linear-polarization values. The theoretical values
of the $R_\textrm{DCO}$ and linear polarization were computed as a function
of mixing ratio. Here, the theoretical contours were obtained by
assuming both the positive and negative parity for the 331-keV state.
The experimental point for the 331-keV $\gamma$ ray is observed to
be equidistant to both these contours (see Fig. \ref{mixRatio_331}). 
The comparison of the experimental data point with the contour for the
\textit{M}1+\textit{E}2 multipolarity suggests a mixing ratio value
of 6.1(36), which corresponds to $\sim$ 97\% \textit{E}2 component
in the 331-keV transition. Another possibility is that the
331-keV $\gamma$ ray is an \textit{E}1+\textit{M}2 transition 
with $\sim$ 94\% \textit{M}2 component [$\delta$ = 3.8(8)]. The
Weisskopf estimate for an \textit{E}1+\textit{M}2 transition
with dominant \textit{M}2 component suggests a half life of
$\sim$ 75 ns for the 331-keV state. However, the \textit{early-delayed}
analysis does not indicate the isomeric nature of the 331-keV
state. Thus, the possibility of the 331-keV $\gamma$ ray being
an \textit{E}1+\textit{M}2 transition with a dominant
\textit{M}2 component can be excluded. The present analysis
suggests a mixed \textit{M}1+\textit{E}2
multipolarity for the 331-keV $\gamma$ ray and consequently,
a spin-parity of 11/2$^{+}$ to the 331-keV state.
The deduced spin-parity of the 331-keV state is consistent with
the previous assignments from $\gamma$-ray spectroscopic
studies as well as $\alpha$-decay studies \cite{217Ra_Roy,217Ra_Sugawara,
217Ra_Sugawara_RIKEN,217Ra_Muralithar,217Ra_AlphaDecay}.

%
\makeatletter
\def\LT@LR@e{\LTleft\z@   \LTright\z@}%
\makeatother
\afterpage{
\onecolumngrid
\begin{center}
\setlength\LTcapwidth{\textwidth}
\renewcommand\thefootnote{\alph{footnote}}
%

\begin{longtable}[H]{@{\extracolsep{\fill}}ccccccccc}
\caption{\label{Table I}
Table of $\gamma$-ray energies, level energies, spin-parity 
of the initial and final states, relative $\gamma$-ray intensities 
of the transitions in $\mathrm{^{217}Ra}$. The $R_\textrm{DCO}$
values of the transitions in dipole and quadrupole gates are
mentioned along with the linear polarization values and
deduced multipolarity of the transitions. 
The reported errors in the $\gamma$-ray energies, 
relative $\gamma$-ray intensities and $R_\textrm{DCO}$ values include 
contribution from both the systematic and statistical uncertainties. 
The systematic uncertainty in $I_\gamma$ is considered to
be 5\% of the relative $\gamma$-ray intensity.}\\
\hline
\hline
\multicolumn{1}{c}{{$E_\gamma$} (keV)} &
\multicolumn{1}{c}{$E_{i}$ (keV)}&
\multicolumn{1}{c}{$I_{i}^\pi$} &
\multicolumn{1}{c}{$I_{f}^\pi$} &
\multicolumn{1}{c}{{$I_\gamma$}} &
\multicolumn{1}{c}{$R_\textrm{DCO}^{Q}$} &
\multicolumn{1}{c}{$R_\textrm{DCO}^{D}$} &
\multicolumn{1}{c}{$\textit{P}$} &
\multicolumn{1}{c}{Multipolarity} \\
\hline
\endfirsthead
\hline
\endhead
\caption[]{(\textit{Continued.})}\\
\hline
\hline
\multicolumn{1}{c}{{$E_\gamma$} (keV)} &
\multicolumn{1}{c}{$E_{i}$ (keV)}&
\multicolumn{1}{c}{$I_{i}^\pi$} &
\multicolumn{1}{c}{$I_{f}^\pi$} &
\multicolumn{1}{c}{{$I_\gamma$}} &
\multicolumn{1}{c}{$R_\textrm{DCO}^{Q}$} &
\multicolumn{1}{c}{$R_\textrm{DCO}^{D}$} &
\multicolumn{1}{c}{\textit{P}} &
\multicolumn{1}{c}{Multipolarity} \\
\hline
\endhead
\hline
\endfoot
\hline \hline
\endlastfoot
  92.9(5) &2394.1(5)  & 33/2$^{-}$ & 29/2$^{-}$ &2.1-3.9\footnotemark[1]& & & & \\
  119.7(3) &1337.5(5) & 19/2$^{+}$ & 17/2$^{-}$ &1.3(3) &0.56(7) &1.16(16) & & \it{D} \\
  123.1(3) &3629.6(4)  & 41/2$^{-}$ & 39/2$^{+}$ & 4.6(2)&0.44(5) & & & \it{D}\\
  (125.0)\footnotemark[2] &3258.2(1)  & 37/2$^{-}$ & 35/2$^{+}$ &- & & & & \\
  127.1(1) &666.7(1)  & 15/2$^{-}$ & 13/2$^{+}$ & 49.9(25)&0.67(4) &1.20(7) & &\textit{E}1\footnotemark[3] \\
  142.4(4) &4328.8(4)  & 49/2 &47/2$^{+}$  &0.9(1) &0.64(3) & & & \it{D}\\
  157.1(2) &1611.7(3)  & 23/2$^{-}$ & 21/2$^{+}$ & 7.4(4)& 0.63(4)&1.15(8) & &\textit{E}1\footnotemark[3] \\
  171.3(2) &1173.5(2)  & 19/2$^{-}$ & 17/2$^{+}$ & 13.4(7)& 0.60(3) &1.07(7) & &\textit{E}1\footnotemark[3] \\
  176.9(4) &5001.4(4)  & 53/2 & 51/2 &0.3(1) &0.49(3) & & & \it{D}\\
  196.7(2) &3826.3(2)  & 45/2$^{-}$ & 41/2$^{-}$ & 6.2(3)& 1.11(9)&2.07(24) & &\textit{E}2\footnotemark[3] \\
  208.9(3) &539.6(3)  & 13/2$^{+}$ & 11/2$^{+}$ & 1.0(1) & & & & \\
  213.1(2) &1667.6(3)  & 23/2$^{+}$ & 21/2$^{+}$ &13.0(7) & 0.49(3)&1.00(6) & &\it{D} \\
  216.7(3) &1884.3(5)  & 25/2 & 23/2$^{+}$  &2.0(1) &0.51(6) &0.94(19) & &\it{D} \\
  229.0(1) &1896.6(1)  & 27/2$^{+}$ & 23/2$^{+}$ &47.8(24) &1.07(6) &1.94(12) &0.75(11) &\textit{E}2 \\
  235.4(4) &931.1(5)  & 15/2$^{+}$ & 13/2$^{-}$ &0.3(1) & &0.99(22) & &\it{D} \\
  248.3(3) &3506.7(4)  & 39/2$^{+}$ & 37/2$^{-}$ & 4.5(2)&0.52(4) &0.98(11) & &\it{D} \\
  271.4(4) &2301.2(5)  & 29/2$^{-}$ & 27/2$^{-}$ &0.9(1) &0.49(5) & & &\it{D} \\
  275.1(3) &2305.3(3) & 31/2 & 27/2$^{-}$ & 1.2(1)&1.00(12) & & &\it{Q} \\
  281.4(1) &1454.8(4)  & 21/2$^{+}$ & 19/2$^{-}$ &30.4(15) &0.55(3) &0.98(6) &0.49(8) &\textit{E}1 \\
  287.4(3) &1218.3(4)  & 17/2$^{-}$ & 15/2$^{+}$ &2.9(2) &0.53(6)&1.08(17) &0.54(11) & \textit{E}1\\
  330.0(1) &1667.6(3)  & 23/2$^{+}$ & 19/2$^{+}$ &25.9(13) &1.06(6) & &0.45(9) &\textit{E}2 \\
  330.7(1) &330.7(1)  & 11/2$^{+}$ & 9/2$^{+}$ &46.7(39) & &1.45(13) &0.10(5) & \textit{E}2(+\textit{M}1)\footnotemark[4]\\
  335.6(2) &1337.5(5)  & 19/2$^{+}$ & 17/2$^{+}$ &8.9(5) &0.54(4) & &-0.10(3) & \textit{M}1\\
  335.7(1) & 1002.2(3) & 17/2$^{+}$ & 15/2$^{-}$ &28.7(25) & &1.00(7) &0.25(5) &\textit{E}1 \\
  344.0(3) &3599.4(3) &41/2  &37/2$^{-}$  &1.4(1) &0.89(10) & & &\it{Q} \\
  349.4(5) &3604.8(5) &39/2  &37/2$^{-}$  &1.3(2) &0.50(7) & & &\it{D} \\
  352.1(4) &3956.9(4) & (41/2) &39/2  &1.1(1) &0.75(14) & & & \it{D+Q}\\
  360.1(3) &4186.4(3)  & 47/2$^{+}$ & 45/2$^{-}$ &3.6(2) &0.54(4) & 0.84(10)&0.37(7) & \textit{E}1\\
  360.2(5) &1971.9(5)  & 25/2$^{(+)}$ & 23/2$^{-}$ &2.2(4) & &0.85(9) & &\it{D} \\
  365.0(2) &695.7(2)  & 13/2$^{-}$ & 11/2$^{+}$ &5.7(3) &0.53(8) & &0.58(12) &\textit{E}1 \\
  371.3(2) &3629.6(4)  & 41/2$^{-}$ & 37/2$^{-}$ &9.4(5) &0.99(7) &1.91(19) &0.64(13) &\textit{E}2 \\
  373.6(3) &3506.7(4)  & 39/2$^{+}$ & 35/2$^{+}$ &2.4(1) &1.08(13) & &0.56(13) &\textit{E}2 \\
  379.5(4) &3182.3(7)  &35/2  & 33/2$^{-}$ &1.0(1) &0.52(7) & & & \it{D}\\
  391.4(3) &931.1(5)  & 15/2$^{+}$  &  13/2$^{+}$ &2.5(1) &0.73(6) & & & \it{D+Q}\\
  395.0(4) &1732.1(6)  & 21/2$^{(-)}$  &  19/2$^{+}$ &1.0(1) &0.47(5) & & & \it{D}\\
  404.6(1) &2301.2(5)  & 29/2$^{-}$  &  27/2$^{+}$ &63.5(32) &0.51(3) &1.02(6) &0.39(8) & \textit{E}1\\
  406.3(1) &1337.5(5)  & 19/2$^{+}$  &  15/2$^{+}$ &25.7(57) &0.85(10) & & & \it{Q}\\
  (409.9)\footnotemark[2] &3133.2(3)  & 35/2$^{+}$  &  31/2 &- & & & & \\
  418.5(1) &2030.2(1)  & 27/2$^{-}$  &  23/2$^{-}$ & 17.0(9)&1.03(6) &1.81(15) &0.57(12) &\textit{E}2 \\
  425.6(3) &3255.4(4)  & 37/2$^{-}$  &  33/2$^{-}$ &4.1(2) &1.08(11) & &0.64(14) &\textit{E}2 \\
  429.6(5) & 1884.3(5) & 25/2  &  21/2$^{+}$ &0.9(2) & 0.86(11)& & & \it{Q}\\
  430.8(5) & 2098.4(5) & 25/2  &  23/2$^{+}$ &1.8(1) &0.46(5) & & & \it{D}\\
  438.1(1) &1611.7(3)  & 23/2$^{-}$  &  19/2$^{-}$ &16.8(13) &0.96(5) & &0.53(12) &\textit{E}2 \\
  438.5(2) & 2832.6(2) & 35/2 $^{(-)}$ &33/2$^{-}$ & 7.6(4) &0.59(4) &1.05(17)&0.03(3) &(\textit{M}1+\textit{E}2) \\
  446.5(4) & 2114.1(4) &  25/2 &  23/2$^{+}$ & 1.0(1)&0.62(7) & & &\it{D} \\
  452.4(1) &1454.8(4)  & 21/2$^{+}$  &  17/2$^{+}$ & 21.7(14)&0.99(8) & &0.48(12) &\textit{E}2 \\
  452.8(2) & 3255.4(4) & 37/2$^{-}$  &  33/2$^{-}$ &5.3(3) &1.15(12) & &0.51(13) & \textit{E}2\\
  462.3(1) &1002.2(3)  & 17/2$^{+}$  &  13/2$^{+}$ &27.1(14) &1.04(8) &1.78(14) &0.65(15) & \textit{E}2\\
  464.9(3) & 2986.6(3) & 33/2  &  31/2$^{-}$ & 3.0(1)&0.49(5) &1.15(15) & & \textit{D}\\
  489.2 (4) &4315.5(4)  &   & 45/2$^{-}$  &0.7(1) & & & & \\
  491.5(3) &2521.7(3)  & 31/2$^{-}$  &  27/2$^{-}$ &3.2(2)&1.08(8) & &0.87(20) &\textit{E}2 \\
  495.7(4) &4824.5(4)  & 51/2  &  49/2 & 0.4(1)&0.47(2) & & &\it{D} \\
  498.6(4) &2528.8(4) &29/2  &  27/2$^{-}$ &0.7(1) &0.48(6) & & & \it{D}\\
  501.4(2) &2802.6(2) & 33/2$^{-}$  &  29/2$^{-}$ &8.4(4) &1.07(8) &2.12(22) &0.85(21) &\textit{E}2 \\
  506.8(1) &1173.5(2) & 19/2$^{-}$  &  15/2$^{-}$ &37.0(19) &1.04(6) &1.82(10) &0.69(17) &\textit{E}2 \\
  513.4(5) &1732.1(6) & 21/2$^{(-)}$  &  17/2$^{-}$ &0.4(1) & & & & \\
  520.8(4) &2132.5(4) & 25/2  &  23/2$^{-}$ &0.7(1) &0.54(3) & & &\it{D} \\
  522.5(3) &1218.3(4) & 17/2$^{-}$  &  13/2$^{-}$ &1.6(2) & &2.06(25) & &\it{Q} \\
  528.7(4) &2829.9(4) & 33/2  &  29/2$^{-}$ &0.8(1) &0.86(13) & & & \it{Q}\\
  539.6(1) &539.6(3) & 13/2$^{+}$  &  9/2$^{+}$ &100.0(52) &0.94(5) &1.86(1) &0.49(13) &\textit{E}2 \\
  566.4(4) &2178.1(4) &   &  23/2$^{-}$ &2.1(1) & & & & \\
  570.3(4) &1266.0(4) &   &  13/2$^{-}$ &0.8(1) & & & & \\
  600.4(1) &931.1(5) &  15/2$^{+}$ &  11/2$^{+}$ &28.2(15) &0.90(5) & & 0.50(14) &\textit{E}2 \\
  637.7(4) &3440.3(4) &   &  33/2$^{-}$ &0.9(1) & & & & \\
  641.8(3) &3163.5(3) & 35/2 &  31/2$^{-}$ &1.9(1) &0.89(5) & & &\it{Q} \\
  648.2(4) &1821.7(4) &21/2 &  19/2$^{-}$ &0.8(1) &0.54(5) & & &\it{D} \\
  670.2(4) &1365.9(4)  & 17/2 &  13/2$^{-}$ &1.0(1) & &1.78(28) & &\it{Q} \\
  739.1(3) &3133.2(3) & 35/2$^{+}$  &  33/2$^{-}$ &3.0(1) &0.51(4)&0.82(9) &0.33(13) &\textit{E}1 \\
  788.2(3) &3182.3(7) &35/2 & 33/2$^{-}$ & 1.1(1)&0.57(3)& & & \it{D}\\
  826.7(3) &2723.3(3) & 31/2  &  27/2$^{+}$ &1.4(1) & 0.81(10)& & & \it{Q}\\
  864.1(1) &3258.2(1) & 37/2$^{-}$  &  33/2$^{-}$ &16.4(8) &1.04(6) &1.77(14) &0.93(41) &\textit{E}2 \\
  881.2(4) &3182.3(7) &35/2 & 29/2$^{-}$ & 0.8(1)& & & & \\
\end{longtable}
\vspace*{-8mm}
  \begin{minipage}{2\textwidth}
  \begin{flushleft}
   \footnotetext[1]{{Intensity range on the basis of total intensity balance at
   the 29/2$^{-}$ and 33/2$^{-}$ states assuming the \textit{E}2 multipolarity of the 93-keV transition.}}
 \footnotetext[2]{{$\gamma$-ray energy from the difference of corresponding level energies;
centroid of the $\gamma$ ray could not be unambiguously determined due to strong close lying transitions.}}  
 \footnotetext[3]{{From intensity balance considerations.}}
 \footnotetext[4]{{$\delta$ = 6.1(3.6); large error in the $\delta$ value can be attributed to the contribution of both
systematic and statistical uncertainties in the $R_\textrm{DCO}$ value.}}

  \end{flushleft}
  \end{minipage}
\end{center}
\twocolumngrid
}


Similarly, the states up to 23/2$^{+}$ were established and the
observed level structure was found to be in agreement with
that reported in the earlier studies. The $\gamma$-$\gamma$
coincidence relationships suggest the presence of 229- and 405-keV
transitions above the 23/2$^{+}$ state. However, their relative
ordering was found to be inconsistent in Refs. \cite{217Ra_Roy}
and \cite{217Ra_Muralithar}. Roy \textit{et al.} \cite{217Ra_Roy}
placed the 229-keV transition below the 405-keV $\gamma$ ray on
the basis of the presence of a prompt component in the time spectrum
of the 229-keV transition. On the other hand, the ordering of these two
transitions was reversed and a 25/2$^{+}$ state at 2.1 MeV energy was
introduced in the level scheme proposed in Ref. \cite{217Ra_Muralithar}.
In our work, a new 827-keV $\gamma$ ray was identified in the gates
of all the transitions below the 1668-keV state. A comparison of the
gates of the 229- and 405-keV transitions, shown in Fig. \ref{405_pl}(a)
and Fig. \ref{405_pl}(b), clearly reflects the coincidence between the
229- and 827-keV transitions. Furthermore, 
the placement of the 827-keV $\gamma$ ray is supported by the presence of the
410- and 374-keV $\gamma$ rays in the gate of the 827-keV transition.
Thus, the relative ordering of the 229- and 405-keV transitions
and a 27/2$^{+}$ state at 1897 keV were confirmed.

The $R_\textrm{DCO}$ values for the 405-keV $\gamma$ ray in the
gate of stretched dipole and quadrupole transitions
[see Table \ref{Table I} and Figs. \ref{405_pl}(c)]
unambiguously suggest its $\Delta I$ = 1 character,
which is also in agreement with the adopted \cite{217Ra_NDS} dipole
nature of the transition based on the earlier angular distribution
measurements \cite{217Ra_Roy,217Ra_Sugawara}.
Therefore, commensurate with the observed intensity of the
405-keV transition and the underlying reaction mechanism, we assign
$I = 29/2$ to the 2301-keV state. Furthermore, the linear-polarization
measurements clearly indicate the electric nature of the transition.
The deduced \textit{E}1 multipolarity of the 405-keV $\gamma$ ray is
in contrast to its previously assigned \textit{M}1+\textit{E}2 character
with the dominant \textit{E}2 component \cite{217Ra_Roy}.
Although the present data unambiguously suggest a $I^{\pi}$ = 29/2$^{-}$
to the 2301-keV state, the resulting variance with the previously
reported assignments entailed us to further confirm the nature of the
405-keV transition using the mixing ratio calculations.
The theoretical $R_\textrm{DCO}$ and linear-polarization values
were computed by assuming both the positive and negative parity
of the 2301-keV state (see Fig. \ref{mixRatio_331}). It is apparent
that the experimental data point is closer to the theoretical contour
corresponding to the \textit{E}1+\textit{M}2 multipolarity and the
point of intersection of the experimental data point on the contour
suggests a near zero $\delta$ value. This further corroborates the pure
\textit{E}1 nature of the 405-keV $\gamma$ ray and a spin-parity of
29/2$^{-}$ for the 2301-keV level.

\begin{figure}[b]
 \begin{center}
  \includegraphics*[width=\linewidth]{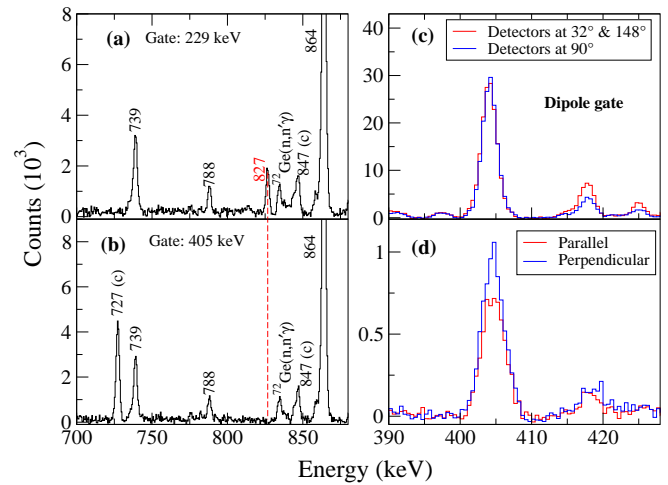}
  
   \caption{\label{405_pl}Evidence for the placement and multipolarity of
   the 405-keV transition. Panels (a) and (b) illustrate the $\gamma$ rays in
   coincidence with the 229- and 405-keV transitions, respectively.
   The presence/absence of the 827-keV $\gamma$ ray is indicated.
   The contaminant peaks marked with (c) are from $\mathrm{^{56}Fe}$
   and $\mathrm{^{212}Po}$. Panel (c) shows a part of the representative
   gated $\gamma$-ray spectrum for the $R_\textrm{DCO}$ measurements of
   the 405-keV $\gamma$ ray and Panel (d) shows the counts in the parallel
   and perpendicular crystals of the $90^{\circ}$ detectors reflecting
   the electric nature of the 405-keV transition.}
 \end{center}
\end{figure}


Figures \ref{g864g739g501}(a) and \ref{g864g739g501}(b) illustrate
the $\gamma$ rays in the gate of 864- and 501-keV transitions,
respectively. It is evident that both these $\gamma$ rays
are in coincidence with all the transitions below the
29/2$^{-}$ state. Therefore, the 864- and 501-keV transitions
can be placed directly above the 29/2$^{-}$ level. However, an
isomeric state [$T_{1/2}$ = 4.62(6) ns] with $I^{\pi}$ = 33/2$^{+}$ was
identified at 2.4 MeV excitation energy and the 864-keV $\gamma$ ray
was placed above the isomer in the previous studies
\cite{217Ra_Roy,217Ra_Sugawara_RIKEN,217Ra_NDS,217Ra_Dracoulis}. Also, it was reported that the
33/2$^{+}$ isomer deexcites via a 93-keV \textit{E}2 transition. 
Figure \ref{g_LEPSvsALL} shows the coincidence spectra observed 
in LEPS gated by the $\gamma$ rays recorded in the clover detectors. 
The presence of the 93-keV $\gamma$ ray in the gate of the 864-keV 
transition [Fig. \ref{g_LEPSvsALL} (a)] is in agreement with the
previous placement of these transitions. Furthermore, in the earlier work,
the 93-keV $\gamma$ ray was assigned \textit{E}2 character on the
basis of conversion-coefficients measurements \cite{217Ra_Roy}.
Considering the observed intensity (see Fig. \ref{g_LEPSvsALL})
and the theoretical conversion-coefficients for the possible
multipolarity (\textit{E}1, \textit{M}1, \textit{E}2, \textit{M}2...)
of the 93-keV $\gamma$ ray, the present data also supports its
\textit{E}2 multipolarity. Since the present measurements have assigned
the $I^{\pi}$ = 29/2$^{-}$ to the 2301-keV state, a spin-parity of
33/2$^{-}$ is deduced for the state at 2394 keV excitation energy.


\begin{figure}[t]
 \begin{center}
  \includegraphics*[width=\linewidth]{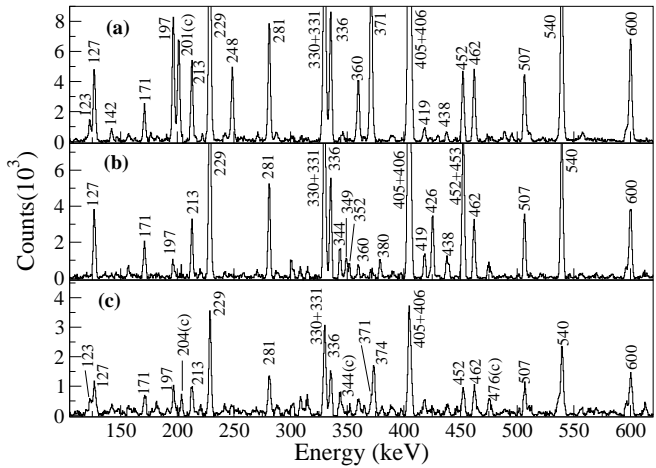}
  
   \caption{\label{g864g739g501} $\gamma$-ray coincidence spectra
   illustrating transitions in the gate of the (a) 864- (b) 501-
   and (c) 739-keV transitions. The contaminant peaks marked with (c)
   are mainly from $\mathrm{^{65}Zn}$ and $\mathrm{^{216}Ra}$.}
 \end{center}
\end{figure} 

\begin{figure}[b]
 \begin{center}
  \includegraphics*[width=\linewidth]{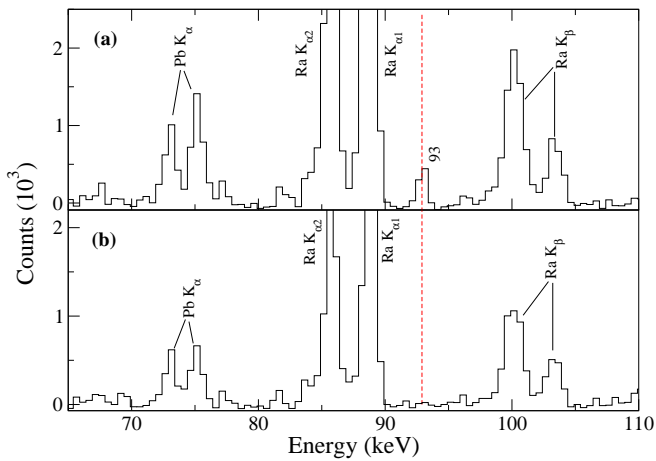}
  
   \caption{\label{g_LEPSvsALL}Coincidence $\gamma$-ray spectra
   observed in the LEPS gated by the (a) 864- and (b) 501-keV
   $\gamma$ rays recorded in the clover detectors.}
 \end{center}
\end{figure}

Furthermore, the half-life of the 2394-keV state has been revisited
using the centroid-shift analysis. In our work, the
cumulative half-life of the 2301- and 2394-keV states was determined
from the time difference between the 864- and 405-keV $\gamma$ rays.
The blue curve in Fig. \ref{C_Shift} shows the time difference between 
the 864- and 405-keV $\gamma$ rays which is compared with the time
difference spectrum (in red color) of prompt transitions with similar energies. 
It is to be noted that the time difference spectra used for the
comparison are generated after taking proper background subtraction
into account. The cumulative half-life of the 2301- and 2394-keV states
is deduced to be 3.9(11) ns which is found to be consistent with
the adopted value of 4.62(6) ns for the 2394 keV state \cite{217Ra_NDS}.
This suggests that the measured cumulative half-life is mainly due to
the half-life of 2394-keV level. The error quoted in the half-life is
obtained by adding the systematic and statistical errors in 
quadrature. The systematic error includes contributions from the discrete
binning along the time axis and the uncertainty arising due to shift
in the centroid of the prompt distribution of two similar energy transitions.

\begin{figure}[t]
 \begin{center}
  \includegraphics*[width=\linewidth]{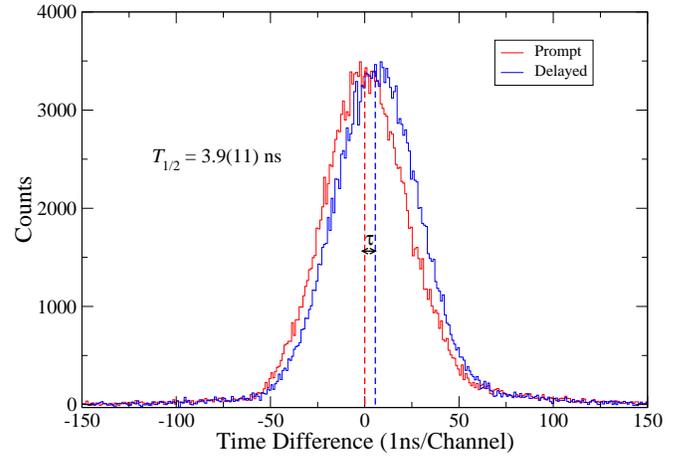}
  
   \caption{\label{C_Shift}
   Centroid-shift analysis to determine
   the cumulative half-life of the 2301- and 2394-keV states.
   The blue curve for the time difference between the 864- and 405-keV 
   $\gamma$ rays is compared with the time difference spectrum
   obtained using prompt transitions of similar energies. The difference
   between the centroids of the prompt and delayed
 distributions gives the cumulative mean life of the states.}
 \end{center}
\end{figure}

Furthermore, the placements of the 501- and 739-keV transitions were
found to be at variance in Refs. \cite{217Ra_Sugawara_RIKEN} and
\cite{217Ra_Muralithar}. The absence of the 93-keV transition in the
gate of the 501-keV $\gamma$ ray [Fig. \ref{g_LEPSvsALL}(b)] confirms
the placement of this transition as demonstrated in the present level
scheme. Similar method could not be adopted to confirm the placement
of the 739-keV $\gamma$ ray owing to its weak intensity. However, the
presence of the 123-, 197- and 374-keV transitions in Fig. \ref{g864g739g501}(c)
confirms the placement of the 739-keV $\gamma$ ray as shown in the
Fig. \ref{LS}. Similar analysis techniques were employed to confirm
the remaining level structures reported in the earlier works.

\subsection{Evidence of parity doublets in $\mathbf{^{217}Ra}$}

Along with the two positive-parity and a 
negative-parity sequences below the isomer, a 1050-keV state
was established on the basis of coincidence between
287- and 119-keV $\gamma$ rays in the earlier studies
\cite{217Ra_Roy,217Ra_Sugawara_RIKEN,217Ra_Muralithar}.
In addition, a 365-keV transition was tentatively placed above the
1050-keV level in the work of Roy \textit{et al.} \cite{217Ra_Roy}.
However, in Ref. \cite{217Ra_Muralithar} the 365-keV transition
was observed to deexcite the proposed 13/2$^+$ state at 696 keV energy.
Figure \ref{g600_g331} illustrates the coincident transitions in the
gates of the 600- and 331-keV transitions. If the placement by
Roy \textit{et al.} \cite{217Ra_Roy} was correct, then the 600-keV
transition should have been in the coincidence with the 365-keV
$\gamma$ ray. However, the absence of the 365-keV $\gamma$ ray
and a newly observed 523-keV transition in Fig. \ref{g600_g331}(a)
clearly suggest that these transitions are
parallel to the 600-keV $\gamma$ ray. 

\begin{figure}[t]
 \begin{center}
  \includegraphics*[width=\linewidth]{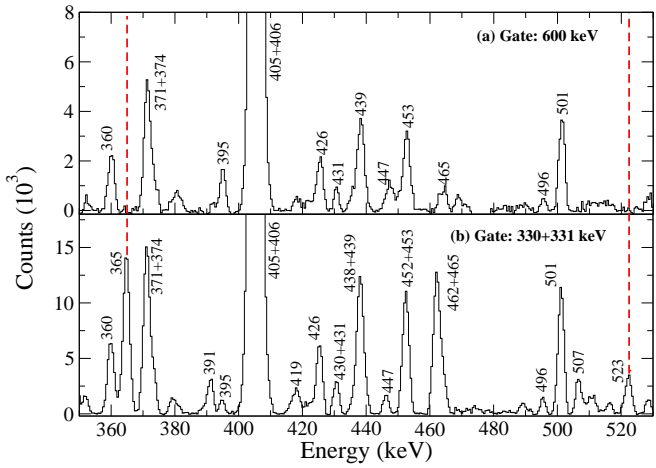}
  
   \caption{\label{g600_g331} A part of the coincidence $\gamma$-ray
   spectra in the gate of the (a) 600-keV transition, and (b) 330-331 keV doublet.
   The presence/absence of the 365-keV transition is indicated.}	
 \end{center}
\end{figure}

\begin{figure}[b]
 \begin{center}
  \includegraphics*[width=\linewidth]{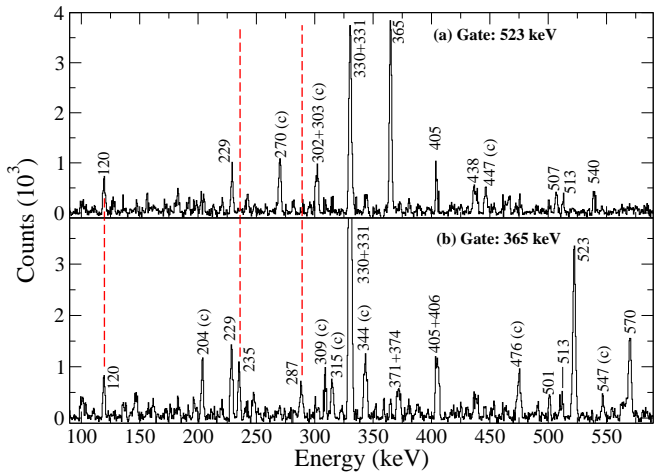}
  
   \caption{\label{g523_g365} A part of the coincidence $\gamma$-ray spectra
   illustrating transitions in the gate of the (a) 523- and (b)
   365-keV $\gamma$ rays provide evidence for reordering the 287- and
   120-keV transitions. The contaminant peaks marked with (c) are mainly
   from $\mathrm{^{214}Rn}$ and $\mathrm{^{216}Ra}$.}
 \end{center}
\end{figure}

Figure \ref{g523_g365} shows the transitions in the gates of the
365- and 523-keV $\gamma$ rays and confirms the coincidence
relationship between these transitions. The placement of the
365- and 523-keV $\gamma$ rays, as indicated in the Fig. \ref{LS}, 
is based on the observed coincidences and intensity considerations. 
Furthermore, the absence of the 287-keV transition in
Fig. \ref{g523_g365}(a) indicates that the ordering of the 287-
and 120-keV transitions as suggested in Refs.
\cite{217Ra_Roy,217Ra_Sugawara_RIKEN,217Ra_Muralithar} should be reversed.
Similar coincidence relationships were utilized to secure
the placement of newly identified 395- and 513-keV transitions
which, in turn, establishes a state at 1732 keV excitation energy.

\begin{figure}[b]
 \begin{center}
  \includegraphics*[width=\linewidth]{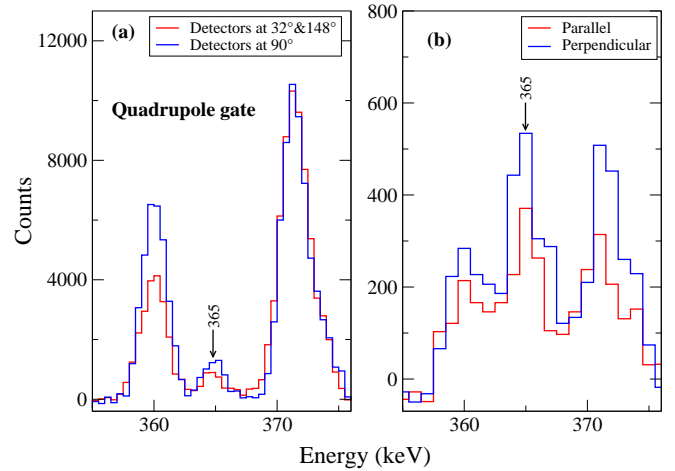}
  
   \caption{\label{365_DCO_PDCO}Representative spectra corresponding
   to the (a) $R_\textrm{DCO}$ and (b) polarization measurements
   of 365-keV transition.}
 \end{center}
\end{figure}

Furthermore, the $R_\textrm{DCO}$ and linear-polarization values
of the observed transitions were utilized in order to assign
the spin-parities to the newly established states.
In the gate of the 229-keV $\gamma$ ray, which is a stretched
$\Delta I$ = 2 transition, the $R_\textrm{DCO}$ value for the
365-keV transition is found to be 0.53(8)
(see Fig. \ref{365_DCO_PDCO}(a) and Table \ref{Table I}).
This suggests the dipole character of the 365-keV
transition. Figure \ref{365_DCO_PDCO}(b) illustrates a part
of the representative spectra for the polarization measurements
and higher counts in the spectrum corresponding to the perpendicular
scatterings with respect to the reaction plane clearly indicates
the electric nature of the 365-keV $\gamma$ ray. Thus, a spin-parity
of 13/2$^{-}$ is assigned to the 696-keV state. Furthermore, the
measured DCO ratios for the 287- and 523-keV transitions and
polarization values for the 287-keV $\gamma$ ray suggest the
spin-parity of 17/2$^{-}$ for the 1218-keV level. 
Similarly, the $R_\textrm{DCO}$ measurements for the 395-keV
transition suggests $I$ = 21/2 for the 1732-keV state. However,
the electric/magnetic character of the transitions depopulating
the 1732-keV state could not be determined due to poor statistics. 
The negative parity of the 21/2 state is tentatively assigned on
the basis of striking similarity in the observed low-lying
structures of $\mathrm{^{217}Ra}$ and $\mathrm{^{219}Th}$ \cite{219Th_Reviol}.
As discussed in Section \ref{Section_1}, parity-doublet 
structures are a signature of octupole correlations in
odd-\textit{A} and odd-odd nuclei. Thus, the newly established
negative-parity sequence provides an experimental evidence of
parity doublets in $\mathrm{^{217}Ra}$ and reflects the expected
similarity in the structures of the transitional
\textit{N} = 129 isotones. Similar analysis techniques were
employed to determine the placement and multipolarity of the
remaining newly identified $\gamma$ rays, wherever possible.

The present data also suggest a few unobserved transitions in
the level scheme based on the $\gamma$-$\gamma$ coincidences.
The coincidence relationships of the 501- and 529-keV $\gamma$
rays with the 426-keV and other higher-lying transitions indicate
the presence of an unobserved 27-keV transition between the 2830-
and 2803-keV states. Similarly, the coincidences of the 197- and
360-keV transitions with the $\gamma$ rays below the 3599-keV state
suggest an unobserved 30-keV transition between the 41/2$^{-}$ state
and 41/2 state at 3599 keV energy. Also, the coincidence 
relationship between the 739- and 371-keV $\gamma$ rays requires a
125-keV transition, which has been reported in earlier studies
\cite{217Ra_Roy, 217Ra_Sugawara_RIKEN}, between the 3258- and
3133-keV states. The centroid and intensity information of the
125-keV transition could not be determined due to
the presence of stronger close-lying 123- and 127-keV transitions. 
Moreover, the 465- and 642-keV transitions are observed to be in
coincidence with the 405-keV $\gamma$ ray and the transitions of
sequence ``C''. This suggests that the 31/2$^{-}$ state deexcites
to the the low-lying  29/2$^{-}$ and/or 33/2$^{-}$ states,
via weak unobserved transitions.
 
Several new transitions (viz. 191-, 242-, 301-, 303-, 736-,
859-, 968 keV), which were assigned to $\mathrm{^{217}Ra}$ in the
present study, could not be unambiguously placed in the level scheme.
These transitions are observed to be in coincidence with all the
$\gamma$ rays below the 29/2$^{-}$ state. However, due to their
weak intensities and/or the lack of firm coincidence relationships 
with the other known $\gamma$ rays above the 29/2$^{-}$ state,
the placement of these transitions could not be ascertained. 
Furthermore, the present data also suggest a doublet character of
the 517-keV transition. One component of the 517-keV doublet is
placed between the 25/2$^{(+)}$ and 21/2$^{+}$ states, whereas the
coincidence relationships suggest the placement of the second
517-keV $\gamma$ ray above the 29/2$^{-}$ level.
In Ref. \cite{217Ra_Sugawara_RIKEN}, the 517-keV transition was
reported to deexcite the 3259-keV state and populate the 2301-keV
level via a 440-keV transition.
However, due to absence of a transition at 440 keV
and other suitable coincidences of the 517-keV transition
above the 29/2$^{-}$ level, the placement of the second
517-keV $\gamma$ ray remains inconclusive in the present
work, similar to the several transitions mentioned above.

\subsection{Intensity Measurements}
The relative intensities of the observed $\gamma$ rays were
determined using the data collected at 72 MeV beam energy. 
The data acquired in the single fold were utilized
to obtain the intensity information for the strong, non-contaminated
(viz. 540-, 229-, 213-, and 462-keV) $\gamma$ rays. The intensities
of all the transitions observed in $\mathrm{^{217}Ra}$ were
normalized with respect to that of the 540-keV $\gamma$ ray,
which is assumed to be 100.
Due to the presence of several doublets in the level scheme,
suitable approaches were employed to determine the relative
intensities of the observed $\gamma$ rays. The 331-keV transition,
which directly feeds the ground state, has also been identified
as a doublet. Therefore, the intensities of the transitions
parallel to the 330-keV (23/2$^{+}\rightarrow19/2^{+}$) $\gamma$
ray were obtained in the efficiency corrected summed gate of
540- and 331-keV transitions and normalized with respect to the
relative intensity of the 213-keV $\gamma$ ray. A few transitions,
which are in a parallel decay path with respect to the
330-keV $\gamma$ ray, such as 452, 438, and 360 keV were
also identified as doublets. Since the doublet counterparts of these
transitions are placed above the isomer and are in coincidence
with both the 330- and 331-keV $\gamma$ rays, their
intensities could not be determined in the summed gate of
the 540- and 331-keV transitions.

In order to determine the intensities of transitions
which are in coincidence with both the $\gamma$ rays of
330-331 keV doublet, it was important to extract the individual
intensity of the 331-keV transition. It may be noted that the
540-keV transition is in coincidence with only the 330-keV
$\gamma$ ray. Therefore, the intensity of the 330-keV transition
was determined in the gate of 540-keV $\gamma$ ray and normalized with
respect to the relative intensity of 462-keV $\gamma$ ray.
The relative intensity thus obtained was further corrected for
the decay branch of the 19/2$^{+}$ state using
the relative intensity of the 395-keV transition.
The total intensity of 330-331 keV doublet was determined from the
single fold data. Thus, the relative intensity of the 331-keV
transition was deduced after subtracting the contribution of the
330-keV $\gamma$ ray from the summed total relative intensity of
the doublet line. Once the relative intensity of the 331-keV
$\gamma$ ray was known, the intensities of the transitions above
the 23/2$^{+}$ state were extracted from the efficiency and branching
corrected 540-keV gate and normalized with respect to the intensity
of the 462-keV $\gamma$ ray. It may be noted that the relative
intensities of the non-contaminated transitions could also 
be determined in the efficiency corrected summed gate of the
540-, 209-, 365- and 600-keV $\gamma$ rays and normalized with
respect to that of the 229-keV transition. 
It was observed that the intensities of the non-contaminated
transitions obtained from the above two approaches were
consistent with each other.

In order to resolve the doublets, wherever possible, appropriate
methods were employed. For example, the total intensity of
the 452-453 keV doublet was determined using the singles data.
The gate of the 281-keV transition, which is in a parallel
decay path with respect to the 452-keV (21/2$^{+}\rightarrow17/2^{+}$)
$\gamma$ ray, was utilized to find the relative intensity of
the 453-keV transition lying above the isomer. The deduced 
intensity was then normalized with respect to the relative intensity
of the 501-keV transition. Further, the intensity of the 452-keV
$\gamma$ ray was extracted by subtracting the contribution of
the 453-keV transition from the doublet line. Similarly, suitable
coincidence relationships were utilized to determine
the intensities of the transitions in the remaining doublets.
  
The above discussion includes the information on the intensity
measurements of all transitions except for the $\gamma$ rays of
the newly established negative-parity sequence. The branching
ratio of the 120- and 406-keV $\gamma$ rays, measured in the
gate of the 395-keV transition, was utilized to deduce the
relative intensity of the 120-keV transition. Furthermore, the
intensity of the 287-keV transition was obtained in the
600-keV gate and normalized with respect to the relative
intensity of the 406-keV transition. The branching of the 
287- and 523-keV $\gamma$ rays in the 120-keV gate was determined
to extract the intensity of the 523-keV transition. The 
relative intensities of the 235-, 570- and 670-keV transitions
were extracted in the gate of the 365-keV $\gamma$ ray and were
normalized with respect to that of the 523-keV transition.
Finally, the intensity of the 365-keV $\gamma$ ray was obtained in 
the gate of the 330-331 keV doublet and the relative intensity
of the 523-keV $\gamma$ ray was used for normalization.

\section{Discussion}
As discussed in Section \ref{Section_1}, $\mathrm{^{217}Ra}$ with
\textit{Z} = 88 and \textit{N} = 129 lies in the transitional
region between the spherical and static octupole-deformed nuclei. 
The earlier reported low-lying sequences starting with 9/2$^{+}$,
11/2$^{+}$ and 15/2$^{-}$ states were qualitatively understood 
in terms of mainly the shell-model configurations viz. $\nu g_{9/2}^{3}$,
$\nu g_{9/2}^{2}i_{11/2}$ and $\nu g_{9/2}^{2}j_{15/2}$,
respectively \cite{217Ra_Roy}. Although these sequences were
reported to terminate at the spins predicted by the above
mentioned configurations, the almost constant level spacings
observed in the ground-state and negative-parity sequences along
with the strong \textit{E}1 connecting transitions reflected the
contribution of collective degrees of freedom in this nucleus.
In the present work, a new negative-parity sequence has been 
established along with the above mentioned three sequences.
The addition of this newly established negative-parity sequence
hints at the parity-doublet structures in $\mathrm{^{217}Ra}$, 
which are expected in a odd-\textit{A} nucleus displaying octupole
correlations. Furthermore, the two alternating-parity sequences
established in $\mathrm{^{217}Ra}$ are characterized by the simplex
quantum number (\textit{s}) \cite{Simplex}. It is known that for 
even-\textit{A} nuclei the sequences 0$^{\pm}$, 1$^{\mp}$, 2$^{\pm}$,
3$^{\mp}$, 4$^{\pm}$,... are associated with the \textit{s = $\pm$}1
quantum number, respectively. In the case of odd-\textit{A} nuclei,
\textit{s = $\pm$i} is assigned to the 1/2$^{\pm}$, 3/2$^{\mp}$,
5/2$^{\pm}$, 7/2$^{\mp}$, 9/2$^{\pm}$,... sequences, respectively.

\begin{figure}[]
 \begin{center}
  \includegraphics*[width=\linewidth]{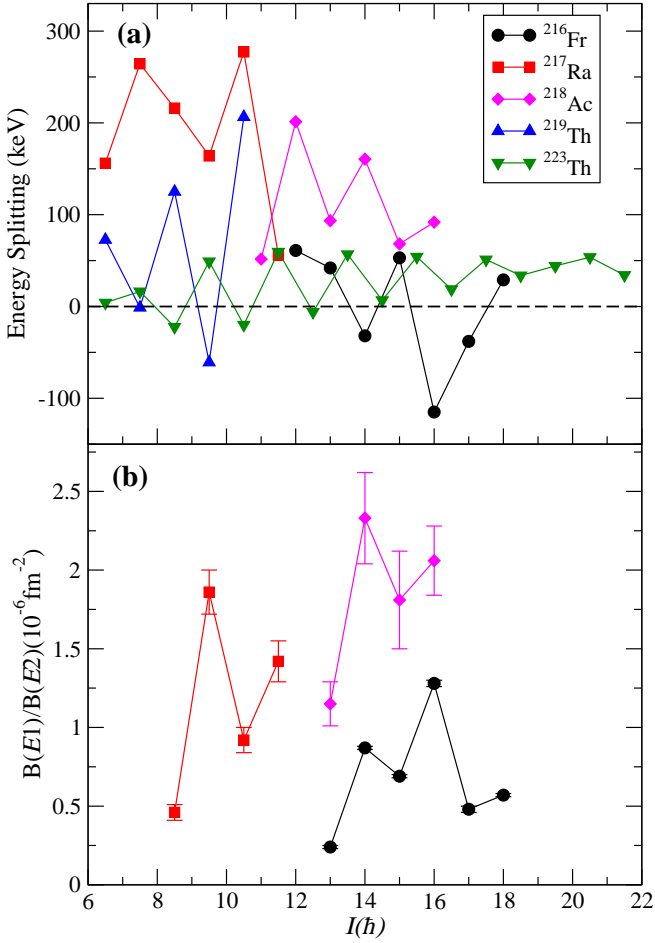}
  
   \caption{\label{ES_BE1byBE2Ratio}Comparison of the (a) energy splitting
[E(\textit{s} = -\textit{i}/-1) - E(\textit{s} = +\textit{i}/+1)]
and (b) B(\textit{E}1)/B(\textit{E}2) ratios versus initial spin for the
transitions of \textit{s} = +\textit{i} band in
$\mathrm{^{217}Ra}$ with \textit{N} = 129 isotones
($\mathrm{^{216}Fr}$, $\mathrm{^{218}Ac}$, $\mathrm{^{219}Th}$).
The data are taken from Refs. \cite{216Fr, 218Ac_Debray, 219Th_Reviol} and the
present work.}
 \end{center}
\end{figure}

\begin{figure}[t]
 \begin{center}
  \includegraphics*[width=\linewidth]{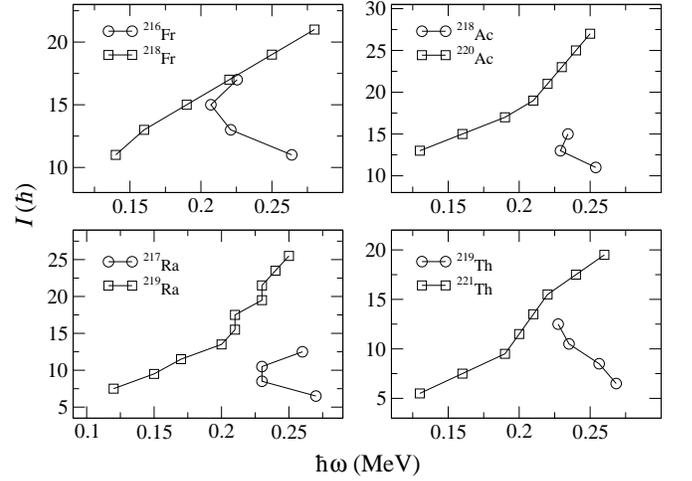}
  
   \caption{\label{SpinVsFreq} Systematics of the spin
   ($I$) versus rotational frequency ($\hbar\omega$) for the lowest lying
   sequence of the \textit{s} = +\textit{i}/+1 bands in the
\textit{N} = 129- and 131 isotones with 87 $\leq Z \leq$ 90
\cite{216Fr, 218Ac_Debray, 219Th_Reviol, 218Fr, 219Ra, 220Ac, 221Th}.}
 \end{center}
\end{figure}

The properties of the observed simplex bands in $\mathrm{^{217}Ra}$
are compared with those of the similar bands in the neighboring nuclei.
The energy splitting, which is defined as the difference between
the excitation energies of the states of same spin but opposite parity/simplex,
indicates the strength of octupole correlations in a given nucleus.
Ideally, a near zero splitting is expected in the nuclei with static
octupole deformation. In the case of $\mathrm{^{223}Th}$
(\textit{Z} = 90 and \textit{N} = 133), an average energy splitting
of $\sim$ 30 keV is observed which suggests a stable deformation 
in this nucleus \cite{223Th}. On the other hand, the large
average energy splitting ($\sim$ 190 keV) in $\mathrm{^{217}Ra}$
[see Fig. \ref{ES_BE1byBE2Ratio}(a)] indicates that the static
octupole deformation has not set in this transitional nucleus.
Figure \ref{SpinVsFreq} illustrates the variation of the initial
spin with rotational frequency for the ${s} = +\textit{i}(+1)$
band in $\mathrm{^{217}Ra}$ and neighboring nuclei. This relation generally
indicates the extent of rotation-like behavior in a nucleus.
It may be noted that the $I(\omega)$ pattern for $\mathrm{^{217}Ra}$ is
consistent with the almost constant level spacings in the
ground-state \textit{E}2 sequence. Also, the similarity
in the $I(\omega)$ trend observed in all the \textit{N} = 129 isotones
suggests that the addition/removal of a proton in the $h_{9/2}$ orbital
does not significantly influence the dependence of excitation energy on
the spin. Furthermore, the comparison of the observed $I(\omega)$ for
the \textit{N} = 129 and \textit{N} = 131 isotones clearly suggests
that the excitation energy of the states better resembles with the
behavior expected for typical rotational bands with the increase in
neutron number. This, in turn, indicates that the quadrupole deformation
in this region is more sensitive to the number of valence neutrons
outside the shell closure.

Furthermore, the experimental B(\textit{E}1)/B(\textit{E}2) ratios 
were extracted using the measured $\gamma$-ray intensities.
The deduced ratios are of the order of $\sim$ 10$^{-6}$ $\mathrm{fm^{-2}}$
which are consistent with the benchmark values if octupole
correlations are present in a nucleus.
It is apparent from Fig. \ref{ES_BE1byBE2Ratio}(b) that the
B(\textit{E}1)/B(\textit{E}2) ratios for the negative-parity states lie
higher than that for the positive-parity states in $\mathrm{^{217}Ra}$,
whereas the converse is true for the odd-odd ($\mathrm{^{216}Fr}$,
$\mathrm{^{218}Ac}$) nuclei. In a stable octupole deformed nucleus, the 
strengths of the electric dipole transitions 
[$I^{\pm} \rightarrow (I-1)^{\mp}$] should be independent of 
the parity of the initial state \cite{Frauendorf_TidalWave}. 
On the other hand, the \textit{E}1 transitions
originating from the positive parity states are forbidden in the 
framework of a pure vibrational model \cite{218Ac_Debray,Frauendorf_TidalWave} .
As the unpaired proton and neutron occupies the $h_{9/2}$ and $g_{9/2}$ 
orbitals, respectively, the higher ratios for the negative (positive)
parity state in odd-\textit{A} (odd-odd) nuclei are consistent with
an intermediate behavior between the extremes discussed above.
It is important to note that the staggering in the B(\textit{E}1)/B(\textit{E}2)
ratios can only be related to the staggering in the B(\textit{E}1)
values if a constant quadrupole deformation is assumed. Therefore,
in order to understand the actual variation of reduced
transition probabilities and their corresponding ratios with angular
momentum, measurements for determining the lifetimes of the states are required.

From the above discussion on the energy splitting, $I(\omega)$, and
B(\textit{E}1)/B(\textit{E}2) ratios, it is evident that the
quadrupole and octupole deformations are not fully developed
in the $\mathrm{^{217}Ra}$. However, the newly established
parity-doublet structures are a clear indication of 
octupole correlations in this nucleus. Therefore, for a more
quantitative understanding, theoretical calculations including
octupole degrees of freedom were performed using reflection-asymmetric
triaxial particle rotor model (RAT-PRM) \cite{RAT-PRM}. 
Recently, RAT-PRM has also been employed to investigate a
possibility of coexistence of nuclear chirality and octupole
correlations in the form of multiple chiral bands
\cite{RAT-PRM_eg1,RAT-PRM_eg2,RAT-PRM_eg3, RAT-PRM_eg4}.

\begin{figure}[b]
\begin{center}
  \includegraphics*[width=\linewidth]{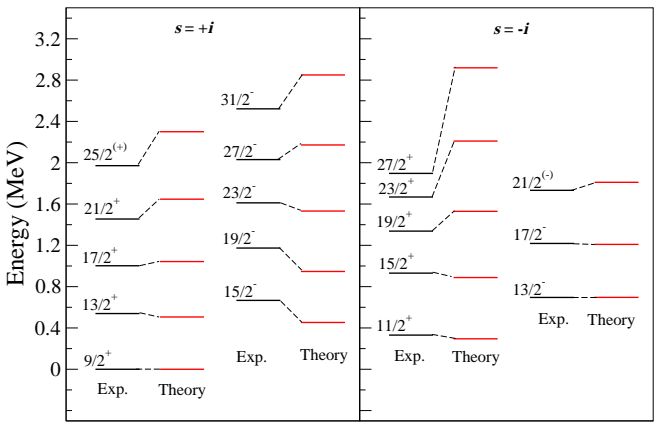}
  
   \caption{\label{Energy_RATPRM_Expt} Comparison of the experimental
   results with the predictions of the RAT-PRM calculations   
   for the simplex bands in $\mathrm{^{217}Ra}$.}
 \end{center}
\end{figure}

The quadrupole and octupole deformation parameters are taken as
input in the RAT-PRM calculations. In the case of $\mathrm{^{217}Ra}$,
the corresponding $\beta_{2}$ = 0.05  and $\beta_{3}$ = 0.08 were
adopted from the Ref. \cite{Moller}. With these deformation parameters, 
the reflection-asymmetric Hamiltonian was solved by expanding the
wavefunction in harmonic oscillator basis \cite{RATPRM_Hamiltonian}.
The single-particle space available to the odd nucleon was truncated
to include six levels, each above and below the Fermi surface.
The pairing correlation was taken into account by an empirical
pairing gap formula $\Delta$ = 12/$\sqrt{A}$ MeV. For the core part,
a spin-dependent moment of inertia $\mathcal{J}_{0}$ = (3.1 + 1.24I) $\hbar^{2}$/MeV
and the core parity-splitting parameter E(0$^{-}$) = 2.0 MeV were used. 
Also, the Coriolis attenuation parameters ($\xi^{\pm}$) for the positive 
and negative parity states were taken to be 0.7 and 1, respectively.
In addition, the electromagnetic transition probabilities
were calculated by assuming intrinsic dipole moment and quadrupole
moments 0.0006*AZe$\beta_{2}\beta_{3}$ and
(3/$\sqrt{5\pi}$)R$_{0}^{2}$Z$\beta_{2}$, respectively \cite{RATPRM_DM, RATPRM_QM}.

The energies calculated from the RAT-PRM calculations for the
\textit{s} = +\textit{i} and \textit{s} = -\textit{i} bands are
compared with the experimental excitation energies in
Fig. \ref{Energy_RATPRM_Expt}. The calculated energies for the
positive-and negative-parity states are shifted to coincide with
the experimental energy of 9/2$^{+}$ and 13/2$^{-}$
states, respectively. It is observed that
the results of the calculations are in good agreement with the
experimental energies up to $I^{\pi}$ = 21/2$^{+}$ for the
positive-parity states and $I^{\pi}$ = 27/2$^{-}$ for the
negative-parity states. The deviation between the calculated and
experimental energies at higher angular momentum may be attributed
to the fact that the vibrational effects could not be included in
the present RAT-PRM calculations.

Since $\mathrm{^{217}Ra}$ is a odd-\textit{N} nucleus, the main
components of the intrinsic wavefunctions of the states are
dominated by the neutron configurations. As the RAT-PRM 
calculations include a finite $\beta_{3}$, the intrinsic parity is
not a good quantum number and the neutron $g_{9/2}$ and $ j_{15/2}$
orbitals mix with each other. The main components of the intrinsic
wave functions were expanded in the strong-coupled basis
$|IMK\rangle$ $\chi^{\nu}$ (denoted as $|K$,$\nu \rangle$
for short), where $|IMK\rangle$ is the Wigner function with
$I$, $M$, and $K$ denoting the quantum numbers of the total
angular momentum and its projections along the third axis in the
laboratory frame and intrinsic frame, respectively, and $\chi^{\nu}$
represents the intrinsic wavefunction of the neutron single-particle
level $|\nu\rangle$. The neutron single-particle levels
$|1\rangle$, $|2\rangle$, $|3\rangle$, $|4\rangle$ and $|5\rangle$
predominately involve $g_{9/2}$ orbital with different
$\Omega$ = 9/2, 7/2, 5/2, 3/2 and 1/2, respectively (see Table \ref{Table II}).

\begin{figure}[b]
\begin{center}
  \includegraphics*[width=\linewidth]{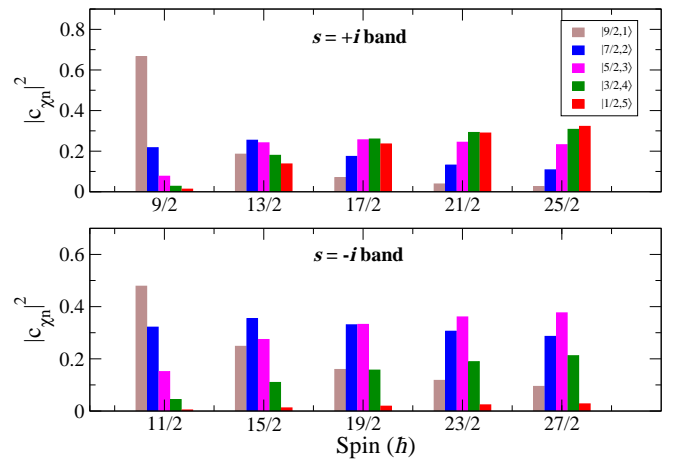}
  
   \caption{\label{WF_PosParity} The probability amplitudes
   of the main components of the intrinsic wave functions
   for the positive-parity states of the $s = \pm i$ bands in $\mathrm{^{217}Ra}$.}
 \end{center}
\end{figure}

Figures \ref{WF_PosParity} and \ref{WF_NegParity} illustrate the main
components of the wavefunctions for the positive- and negative-parity
states of the $s = \pm i$ bands, respectively. It may be noted that
the components corresponding to different neutron single-particle levels
strongly mix with each other. Furthermore, it is known that the
strong Coriolis effects for the low $\Omega$ orbitals result
in lowering the energies of the states which have dominant contribution
from such orbitals \cite{223Th_Calc}. A comparison of the relative
contribution of the single-particle levels with different $\Omega$
values in the wavefunctions of the positive-parity states
(see Fig. \ref{WF_PosParity}) suggests that the states in
$\textit{s}$ = +$\textit{i}$ band are more favored than the corresponding
states in $\textit{s}$ = -$\textit{i}$ band. This is in agreement with
the experimental observations and is reflected in the staggering parameters
[S($I$) = [E($I$)-E($I-1$})]/2$I$] of the positive-parity states
in Fig. \ref{RATPRM_Expt}(a). Similarly, it is observed that the
single-particle level $|5\rangle$ ($\Omega$ = 1/2) significantly
contributes to the wavefunction of all the negative-parity states
of $\textit{s}$ = +$\textit{i}$ band. On the other hand, the
negative-parity states of the $\textit{s}$ = -$\textit{i}$ band
have a dominant contribution from the $\Omega$ = 5/2 orbital. The
effect of the Coriolis interactions suggests the lowering of energies
for the states of the $\textit{s}$ = +$\textit{i}$ band, which also 
agrees with the observed results. The large energy staggering for
the negative-parity states is in agreement with the comparatively
weak intensity of the newly identified negative-parity sequence.

 \setlength{\tabcolsep}{12pt}
\begin{table}
\caption{The main components $l_{j\Omega}$ of the neutron single-particle
levels $|\nu\rangle$ with $\nu$ = 1, 2, 3, 4 and 5.}
\label{Table II}       
\begin{tabular}{cc}
\hline
\hline\noalign{\smallskip}
$|\nu\rangle$ & Main components of the neutron single-particle levels \\
\noalign{\smallskip}\hline\noalign{\smallskip}
$|1\rangle$& 0.948 $g_{9/2,9/2}$ + 0.022 $h_{11/2,9/2}$ + 0.022 $j_{15/2,9/2}$ + ..\\
$|2\rangle$& 0.902 $g_{9/2,7/2}$ + 0.055 $j_{15/2,7/2}$ + 0.018 $f_{7/2,7/2}$ + ..\\
$|3\rangle$& 0.878 $g_{9/2,5/2}$ + 0.076 $j_{15/2,5/2}$ + 0.010 $i_{13/2,5/2}$ + ..\\
$|4\rangle$& 0.810 $g_{9/2,3/2}$ + 0.087 $j_{15/2,3/2}$ + 0.031 $p_{3/2,3/2}$ + ..\\
$|5\rangle$& 0.755 $g_{9/2,1/2}$ + 0.093 $j_{15/2,1/2}$ + 0.067 $p_{3/2,1/2}$ + ..\\ 
\noalign{\smallskip}\hline
\end{tabular}
\end{table}

\begin{figure}[b!]
\begin{center}
  \includegraphics*[width=\linewidth]{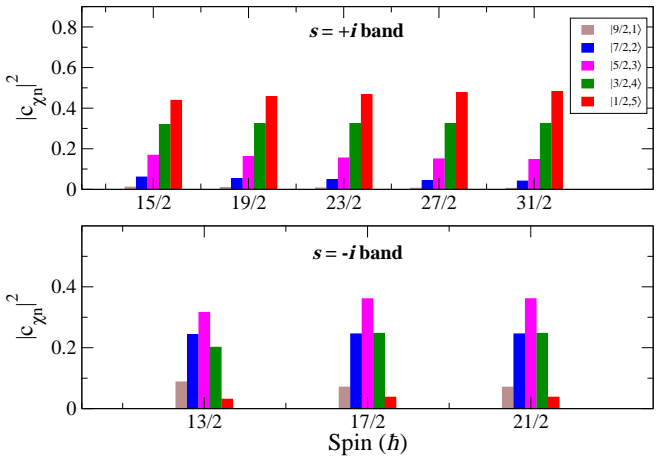}
  
   \caption{\label{WF_NegParity} Same as Fig.
   \ref{WF_PosParity}, but for the negative-parity states.}
 \end{center}
\end{figure}

Figure \ref{RATPRM_Expt} shows a comparison of the observed features of the
simplex partner bands in $\mathrm{^{217}Ra}$ with the results of
the RAT-PRM calculations (dashed lines). It is evident from 
Fig. \ref{RATPRM_Expt} (a) that the RAT-PRM calculations are able
to reproduce the phase of staggering for both the positive- and
negative-parity states. Also, the calculated values are in excellent
agreement with the observed staggering for the negative-parity states,
whereas a good agreement is observed for the positive-parity states
in the spin range 11/2 $\hbar \leq I \leq 21/2$ $\hbar$. 
Figure \ref{RATPRM_Expt}(b) illustrates the spin ($I$) 
vs. rotational frequency ($\hbar\omega$) for the \textit{s} = +$\textit{i}$
band. The experimental $I(\omega)$ relation for both the positive-
and negative-parity states reflects the almost constant level spacing
in these sequences. However, the calculated values exhibit a gradually
increasing spacing between the states of the bands, which would
have been observed if a more pronounced quadrupole deformation 
was present in the nucleus. Thus, the comparison of the
experimental results with the calculated values may suggest that
the adopted quadrupole deformation in the calculations is higher
than the actual deformation. Figure \ref{RATPRM_Expt}(c)
shows the variation of the B(\textit{E}1)/B(\textit{E}2) ratios
with spin for both the \textit{s} = $\pm\textit{i}$ bands and the
calculated results are shown for the comparison. It may be noted that 
the calculated values are almost constant and significantly
underestimated as compared to the observed ratios for both the 
simplex bands. The absence of spin-dependent staggering in the
calculated ratios may be due to the fact that the vibrational
effects could not be taken into account in the calculations.
Furthermore, the calculated and experimental energy splittings between
the states of simplex bands are compared in Fig. \ref{RATPRM_Expt}(d).
It is observed that the calculated energy splittings for the parity
doublets with $I$ = 19/2 $\hbar$ and 23/2 $\hbar$ deviate considerably
from the measured ones. This is in accordance with the large difference
in the calculated excitation energies for the corresponding states
 of the \textit{s} = $\pm\textit{i}$ bands (see Fig. \ref{Energy_RATPRM_Expt}).
Finally, the comparison of the experimental results with the RAT-PRM
predictions further corroborates the fact that the quadrupole and octupole 
deformation are not fully developed in $\mathrm{^{217}Ra}$ and this
transitional nucleus lies between the extremes of spherical and
pear shaped nuclei. Also, further improvements to include the
vibrational effects in the calculations are needed.
\vspace{1cm}

\begin{figure}[t!]
\begin{center}
  \includegraphics*[width=\linewidth]{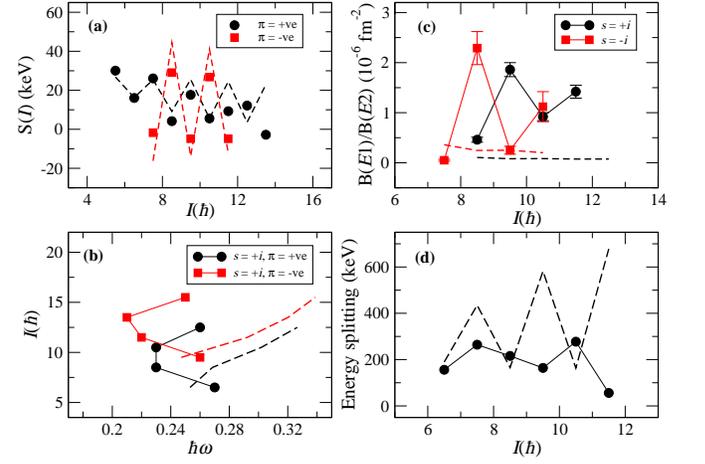}
  
   \caption{\label{RATPRM_Expt} Comparison of the calculated (a) energy
   staggering parameters [S($I$)] (b)  $I$ vs $\hbar\omega$ 
   (c) B(\textit{E}1)/B(\textit{E}2) ratios (d) energy splitting
   with the experimental results for the simplex bands in $\mathrm{^{217}Ra}$.
   The results of the calculations are shown with the dashed lines of the same color.}
 \end{center}
\end{figure}

It is interesting to note that the simplex bands observed in
$\mathrm{^{217}Ra}$ may also be qualitatively interpreted in 
terms of the weak coupling of the odd nucleon to the neighboring
even-even core nuclei. As discussed earlier, the observed level
structures in $\mathrm{^{216}Ra}$ and $\mathrm{^{218}Ra}$ exhibit
the dominance of single-particle and octupole degrees of freedom,
respectively \cite{216Ra_Lonnorth, 218Ra_Schulz}. The level spacings
in the ground-state positive-parity sequence ``A'' of
$\mathrm{^{217}Ra}$ exhibit a vibrational like structure which
is similar to that observed in $\mathrm{^{218}Ra}$, whereas the
decreasing level spacings in the second positive-parity sequence
``C'' suggests the coupling of the odd neutron
in the $i_{11/2}$ orbital to the low-lying (0$^{+}$ - 8$^{+}$) 
states of $\mathrm{^{216}Ra}$. The corresponding negative-parity
sequences may be understood as arising from the coupling of an
octupole phonon to the states of these positive-parity sequences.
Similar interpretation has also been employed in understanding
the observed structures in $\mathrm{^{219}Th}$ nucleus \cite{219Th_Reviol}.

Furthermore, it has been pointed out that the negative-parity
states in the even-even Ra isotopes (\textit{N} $\geq$ 126) reflect
how the structure evolves from spherical to octupole deformed regime
\cite{218Ra220Ra_1}. The lowest negative-parity state ($I^{\pi} = 11^{-}$)
in the lighter Ra isotopes (viz. $\mathrm{^{214}Ra}$ and $\mathrm{^{216}Ra}$)
has been identified above 2.3 MeV excitation energy and interpreted
as two-quasiparticle proton state arising from the
$\pi (h_{9/2}^{5} i_{13/2})$ configuration \cite{216Ra_Lonnorth, 214Ra}.
On the other hand, the presence of an odd-spin negative-parity
\textit{E}2 sequence, interleaved with the ground state positive-parity band,
in the heavier isotopes (viz. $\mathrm{^{218}Ra}$, $\mathrm{^{220}Ra}$)
suggests the onset of octupole collectivity in these nuclei.
In $\mathrm{^{218}Ra}$, a sharp discontinuity at $I^{\pi} = 11^{-}$
($E_{x}$ = 2.1 MeV) is observed in the alignment plot for the
negative-parity states \cite{218Ra220Ra_1}. The sudden increase in
the alignment above the 11$^{-}$ state indicates that a band based on
the two-quasiparticle proton state (11$^{-}$), observed in the lighter
isotopes, crosses the yrast negative-parity band originating from
octupole vibration. Also, the observed decrease in the
B(\textit{E}1)/B(\textit{E}2) ratios above this state provides
further evidence of non-collective effects in the higher-lying 
levels \cite{218Ra220Ra_1,218Ra220Ra_2}. It may be noted that no
features of band crossing at $I^{\pi}$ = 11$^{-}$ were reported 
in $\mathrm{^{220}Ra}$ \cite{218Ra220Ra_1,218Ra220Ra_2}. These
observations are consistent with the expected dominance of the
collective effects over the single-particle degrees of freedom with
the increase in the valence nucleons. Therefore, it may be suggested
that the structure above the 11$^{-}$ and analogous states in the
even- and odd-\textit{A} transitional nuclei, respectively, provide
information on the development of octupole collectivity in this
region. In $\mathrm{^{217}Ra}$, the alternating-parity 
$s$ = -$i$ band is observed to terminate at the 27/2$^{+}$ state
and the level structure above this state seems to be dominated
by the single-particle excitations. It is interesting to note
that a 29/2$^{-}$ state in $\mathrm{^{217}Ra}$ is observed at the 
similar excitation energies as the 11$^{-}$ states in 
$\mathrm{^{216}Ra}$ and $\mathrm{^{218}Ra}$. Therefore, the 29/2$^{-}$
state may be thought of as originating from the coupling of 
odd particle (hole) in the $g_{9/2}$ orbital to the 
11$^{-}$ states in $\mathrm{^{216}Ra}$ ($\mathrm{^{218}Ra}$).
Furthermore, the observed structural change in $\mathrm{^{217}Ra}$
above the 29/2$^{-}$ level also corroborates the negative parity
of the 2301-keV state.
 
The 11$^{-}$ state in $\mathrm{^{216}Ra}$ has the
$\pi (h_{9/2}^{5} i_{13/2}) \otimes \nu (g_{9/2}^{2})$ configuration
\cite{216Ra_Lonnorth,217Ra_Dracoulis}. The addition of an extra 
neutron in the $g_{9/2}$ orbital results in the
$\pi (h_{9/2}^{5} i_{13/2}) \otimes \nu (g_{9/2}^{3})$ configuration
for the 29/2$^{-}$ state. It may further be noted that the
13$^{-}$ state in $\mathrm{^{216}Ra}$ [$T_{1/2}$ = 0.96(20) ns]
decays via an \textit{E}2 transition of 344 keV to the 11$^{-}$ level
\cite{216Ra_Muralithar,217Ra_Dracoulis}. Therefore, the
33/2$_{1}^{-}$ state in $\mathrm{^{217}Ra}$ at 2394 keV
may be realized from the coupling of the odd neutron in the
$g_{9/2}$ orbital to the 13$^{-}$ state in $\mathrm{^{216}Ra}$. 
The coupling of an extra $g_{9/2}$ neutron to the dominant
configuration of the 13$^{-}$ level in $\mathrm{^{216}Ra}$
suggests the $\pi (h_{9/2}^{6}) \otimes \nu (g_{9/2}i_{11/2}j_{15/2})$
configuration for the 33/2$^{-}$ state.

Another possibility is that the 29/2$^{-}$ and 33/2$^{-}$ states in
$\mathrm{^{217}Ra}$ originate from the coupling of odd neutron in
the $j_{15/2}$ orbital to the 8$^{+}$ and 10$^{+}$ states in
$\mathrm{^{216}Ra}$, respectively. The 8$^{+}$ and 10$^{+}$ states
in $\mathrm{^{216}Ra}$ are associated with the
$\pi (h_{9/2}^{6}) \otimes \nu (g_{9/2}^{2})$ and
$\pi (h_{9/2}^{6}) \otimes \nu (g_{9/2}i_{11/2})$
configurations, respectively \cite{216Ra_Lonnorth, 216Ra_Muralithar}.
Thus, the addition of the extra neutron in the $j_{15/2}$ orbital
results in the $\pi (h_{9/2}^{6}) \otimes \nu (g_{9/2}^{2}j_{15/2})$ and
$\pi (h_{9/2}^{6}) \otimes \nu (g_{9/2}i_{11/2}j_{15/2})$ configurations
for the 29/2$^{-}$ and 33/2$^{-}$ states in $\mathrm{^{217}Ra}$,
respectively. Similarly, the coupling of the odd neutron in the
$i_{11/2}$ orbital to the 11$^{-}$ state in $\mathrm{^{216}Ra}$
also results in a 33/2$^{-}$ state in $\mathrm{^{217}Ra}$ with the configuration 
$\pi (h_{9/2}^{5} i_{13/2}) \otimes \nu (g_{9/2}^{2}i_{11/2})$.
The sequence of the 864-, 371- and 197 keV transitions above the isomer
exhibits the similar trend of decreasing level spacings as observed
in the low-lying sequence based on the 11/2$^{+}$ state.  Therefore, the
states of this sequence may be understood as originating from the coupling of the
$[\pi (h_{9/2}^{5} i_{13/2})_{11^{-}} \otimes \nu (g_{9/2}^{2}i_{11/2})_{11/2^{+}}]_{33/2^{-}}$
configuration to the $\nu (g_{9/2}^{2})_{2^{+}-6^{+}}$ neutron excitations.

 \vspace{-1cm}
 \section{Summary}
The level structure of the transitional nucleus $\mathrm{^{217}Ra}$
has been investigated using the $\mathrm{^{208}Pb(^{12}C, 3n){^{217}Ra}}$
reaction. The previously reported level scheme is revisited
and the placements of several $\gamma$ rays, which were found to be 
inconsistent in the earlier studies, have been confirmed in the
present work. The addition of a new negative-parity sequence at
low excitation energies provides the experimental evidence of
parity-doublet structures in $\mathrm{^{217}Ra}$, which were
earlier reported in all the other \textit{N} = 129 isotones with
87 $\leq Z \leq$ 90. Although the level spacings in the \textit{E}2
sequences of the simplex bands do not follow the characteristics of 
typical rotational bands, the strong \textit{E}1 connecting transitions
and parity doublets suggest the presence of octupole correlations
in this nucleus. Therefore, theoretical calculations based on the
reflection-asymmetric triaxial particle rotor model have been performed
to interpret the observed parity-doublet structures. The calculated 
energies were found to be in reasonable agreement with the experimental 
results in the spin range 9/2 $\hbar \leq I \leq 21/2$ $\hbar$ for the
positive-parity states and 13/2 $\hbar \leq I \leq 27/2$ $\hbar$ for
the negative-parity states. The properties (energy splitting, $I(\omega)$
and B(\textit{E}1)/B(\textit{E}2) ratios) of the simplex bands in
$\mathrm{^{217}Ra}$ are also compared with that of the similar bands
in neighboring nuclei and the results of the RAT-PRM calculations. 
The observed structures at low excitation energies reflect an intermediate
behavior between the extremes of single-particle excitations
and stable octupole deformation. Furthermore, it is observed that
the simplex bands in $\mathrm{^{217}Ra}$ terminate at the intermediate
excitation energies, unlike $\mathrm{^{218}Ra}$ where the
alternating-parity sequence persist up to the maximum observed spin. 
Also, the level structures above the 29/2$^{-}$ state in
$\mathrm{^{217}Ra}$ appear to be mainly governed by single-particle
excitations. Finally, it may be noted that the competing
single-particle and collective degrees of freedom govern the structure
in different energy regimes of this nucleus.
  
\section{Acknowledgments}

The authors would like to thank the staff of IUAC target laboratory
for facilitating the preparation of the target.
The authors gratefully acknowledge the IUAC support staff
for their assistance during the experiment. 
The financial support by DST, India (Grant no. IR/S2/PF-03/2003-III) for
the INGA project is acknowledged.
The authors would also like to thank INGA collaboration for their help
and support. Madhu acknowledges the financial support
from DST, India under the INSPIRE fellowship scheme (IF 180082).
AYD and DS would like to acknowledge the financial support by SERB (DST)
vide grant no. CRG/2020/002169.
KY would like to acknowledge the support from the Ministry
of Human Resource Development, India.
Y. Y. Wang, Y. K. Wang and J. Meng would like to acknowledge
the support from the National Natural Science Foundation
of China (Grants No. 12141501, No. 12205097, 
and No. 11935003), and the High-performance Computing Platform
of Peking University.

\end{document}